\documentclass[pdf]{iucr}
\journalcode{D}

\usepackage{float}
\usepackage{amssymb}

\begin{document}

\title{Probing the statistics of sequence-dependent DNA conformations in
  solution using SAXS }
\shorttitle{SAXS study of sequence dependent conformations of DNA}

\cauthor[a]{Heidar J.}{Koning}{heidar.koning@research.uwa.edu.au}{}
\aufn{heidar.koning@research.uwa.edu.au ORCID: 0000-0001-8290-0499}

\author[a]{Anuradha}{Pullakhandam}%
\aufn{anuradha.pullakhandam@research.uwa.edu.au ORCID: 0000-0002-2709-2545}

\author[b]{Andrew E.}{Whitten}%
\aufn{awh@ansto.gov.au ORCID: 0000-0001-8856-3120}

\author[a]{Charles S.}{Bond}%
\aufn{charles.bond@uwa.edu.au ORCID: 0000-0002-9584-6783}

\author[c]{Michel}{Peyrard}%
\aufn{michel.peyrard@ens-lyon.fr ORCID:0000-0002-4047-1363}

\aff[a]{{School of Molecular Sciences},
  {The University of Western Australia},
  {35 Stirling Highway, Crawley}, {WA 6009},
  {Western Australia}, \country{Australia}}

\aff[b]{{Australian Nuclear Science and Technology Organisation (ANSTO)},
  {New Illawarra Rd, Lucas Heights},
  {New South Wales 2234},\country{Australia}}

\aff[c]{{Laboratoire de Physique CNRS UMR 5672},
  {Ecole Normale Sup{\'e}rieure de Lyon},
  {46 all{\'e}e d'Italie, 69007 Lyon}, \country{France}}

\maketitle

\begin{synopsis}
This paper demonstrates how SAXS data on different dsDNA oligonucleotides can
be analyzed using a polymer model to determine the statistical properties of
the local conformations in solution. Orienting the conformations produced by
the model with respect to the sequence allows for the precise determination of
sequence-dependent structural effects.
\end{synopsis}

\begin{abstract}
  SAXS studies of four 60 base-pair DNA duplexes with sequences closely
  related to part of the GAGE6 (G-antigen 6) promoter have been performed to
  study the role of DNA conformations in solution and their potential
  relationship to DNA-protein binding. We show that the SAXS data can be
  analysed using a simple polymer model which nevertheless quantitatively
  describes the average persistence length and torsional rigidity of the DNA
  double helix to determine the statistical distribution of local
  conformations of the DNA in solution to a high accuracy. Although the SAXS
  data is averaged over time and all spatial orientations of the molecules,
  for sequences which have some asymmetry in the data we show that the
  conformations can be oriented with respect to the sequence. This allows
  specific features detected by the analysis to be precisely related to the
  DNA sequence, opening up new opportunities for SAXS to investigate the
  properties of DNA in solution. The biological implications of these results
  are discussed.

  \vspace{1cm}
  \null
\end{abstract}

\section{Introduction}

\label{sec:intro}

In cells, DNA interacts with complex auxiliary machinery which allows its
compact packing \cite{LOWARY}, the reading of the genetic code
\cite{HEUMANN} or the control of gene expression \cite{KNOTT}.
Establishing extensive contacts between a protein and a rigid double
helix is an improbable event. To mitigate this these interactions commonly
require the bending of DNA as typified by DNA wrapping around nucleosomes
\cite{LOWARY} but also when RNA polymerase binds to the double helix
\cite{HEUMANN}. To maximise the number of interactions between a
protein and DNA, the DNA must be able to form surfaces complementary to
binding regions or pockets in proteins. Thus, DNA curvature, flexibility, and
plasticity play an important role, particularly in the selection of a
preferred protein binding site \cite{LOWARY,ANSELMI}.
This has led to many high-resolution structural investigations of free
DNA and DNA-protein complexes \cite{CROTHERS}. Since the pioneering work of
Dickerson and Drew \cite{DICKERSON}, which exhibited the intrinsic
curvature of a DNA dodecamer, many sequences with intrinsic curvature or twist
have been found and tabulated \cite{OLSON1998}, leading to the idea of a
“spatial code hidden within the double helix” \cite{GORIN}.

\medskip

However, crystallographic and cryoEM studies have limitations because they do
not reflect the dynamics of the DNA molecule in solution. DNA’s stiffness does
not prevent it from showing large-scale fluctuations or local fluctuational
openings which may strongly affect its local flexibility
\cite{DNAFLEX}.
These structural dynamics are important both at the global and
local levels and may provide molecular signals for protein binding which are
not apparent in crystallographic and cryoEM data \cite{OLSON1998}.
Molecular dynamic (MD) simulations have been used to investigate the
effects of DNA sequence on its flexibility. MD has allowed the detailed study
of local distortions associated with bending at the scale of a few base pairs
\cite{LANKAS} and may provide insight into the local and global
effects of strong bending \cite{CURUKSU}. These prior studies
establish a theoretical framework for understanding the conformational
dynamics of dsDNA. However, validating these theoretical structural mechanisms
requires that they explain a wide range of experimental observations from a
wide range of structural and biophysical techniques.

\begin{figure*}
  \begin{center}
    \includegraphics[width=14cm,clip]{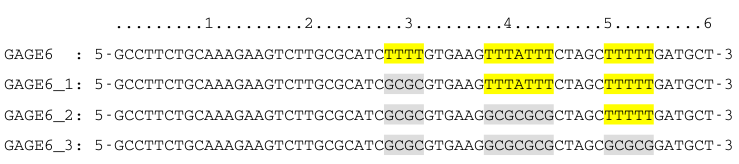}
  \end{center}
  
    {Figure 1:
      DNA sequences. The domains with at least four consecutive
    AT pairs in the GAGE6 sequence are
    highlighted in yellow.
    The grey rectangles mark those domains in which AT pairs have been
    replaced by GC pairs in the sequences GAGE6\_1, GAGE6\_2, GAGE6\_3. 
    The top line marks ensembles of 10 base pairs in the sequences.
    }
  \label{fig:sequences}
\end{figure*}

\medskip
Collecting and analysing data on DNA shape fluctuations in solution is not
easy and is often done via indirect approaches, such as gel migration or
cyclisation kinetics; techniques that lack the spatial resolution to study the
effect of the sequence. Moreover, gels apply external constraints on the
molecules and cyclisation requires peculiar sequences or specific molecular
constructs. Light scattering studies provided the first measurements of the
persistence length of DNA molecules that are unconstrained in solution
\cite{PETERLIN}. Such experiments used visible light and did not have
sufficient spatial resolution to investigate the properties of DNA on the
scale of a few base pairs. However, small-angle X-ray scattering (SAXS) offers
a drastic increase in the resolution of experiments relative to light
scattering. Despite this, extracting the high-resolution properties of DNA in
solution from SAXS data is not straightforward, and this is one aspect that we
discuss herein.

\medskip

A limitation of previous crystallographic and cryoEM studies focusing on DNA
is that they only capture short-range effects. These studies often tabulate
the properties of dimers of base pairs \cite{OLSON1998}, or even up to
tetranucleotides \cite{LAVERY2010} but, as we show in this work, the
change in the mechanical response of a DNA molecule in which the sequence is
locally modified may be more global and extend well beyond a few base
pairs. Being able to detect the most likely conformations of a molecule of a
few tens of base pairs in solution with a sufficient resolution is an
important basis for the discussion of protein-DNA interactions with specific
sequences.

\medskip
In this work, we study a series of four DNA sequences closely related to the
G-antigen 6 (GAGE6) oligonucleotide which is a 60bp dsDNA oligonucleotide that
was first isolated by Song et al. \cite{SONG} as a fragment of the 2241bp
{\em GAGE6}
promoter. This 60bp region was the only part of the digested promoter which
bound the multi-functional nuclear protein Ptb-associated Splicing Factor
(PSF) more commonly known as Splicing Factor Proline/Glutamine rich (SFPQ)
\cite{SONG}. The DNA recognition mode of SFPQ is currently unknown,
although the arginine-glycine (RGG) rich region of the N-terminal
intrinsically disordered region has been identified as the DNA-binding domain
\cite{URBAN,LEE,WANG2022}. The stringent
binding of SFPQ to the {\em GAGE6} promoter,
perhaps suggests elements of the DNA
sequence create a specific binding site for SFPQ. 

\medskip
The native {\em GAGE6} oligonucleotide sequence contains three AT-rich tracts of
varying lengths. AT-rich tracts, in general,  have been the subject of
multiple studies focused on investigating whether they introduce additional
flexibility to the duplex or are energetically capable of causing it to kink
and bend in solution in the absence of protein binding
\cite{CHIRICO2001,HARAN2009,HARTEIS2014,HIZVER2001,MPKINK,STEFL2004,WIDOM1984}.
It is possible with
the GAGE6 fragment that AT-rich regions trigger fluctuational opening of the
base pairs or meta-stable kinks and bends to form in the duplex. These factors
could in turn facilitate enhanced base access for SFPQ. Direct access to bases
in both DNA or RNA is an important feature for RGG-containing binding domains
and their interactions with nucleic acids \cite{CHONG}. To probe the
effect of AT-rich sequences on the GAGE6 oligonucleotide, and to test the
hypothesis that AT-rich domains create fluctuational opening of the base pairs
and localised bending, we synthesised four 60bp dsDNA oligonucleotides where
the AT tracts in the native {\em GAGE6} sequence are incrementally swapped for
GC-rich tracts (termed GAGE6, GAGE6\_1, GAGE6\_2, GAGE6\_3;
Figure 1). These were
then studied using Size Exclusion Chromatography coupled with Small Angle
X-ray Scattering (SEC-SAXS).

\medskip
In this work we show, using the four GAGE6 variants that: 

\begin{itemize}
\item[(i)] the local conformation (bending, twist) of a 60-base pair DNA
  sequence in solution can be derived from the analysis of SAXS data using a
  simplified polymer model, which quantitatively describes both DNA
  persistence length and torsional rigidity. This model allows a very broad
  exploration of the conformational space of the sample and is able to detect
  subtle effects with a resolution on the scale of a few base pairs; 

\item[(ii)] the effect of the sequence on the conformation of dsDNA in
  solution is non-trivial and may not reduce to a sum of local
  effects. Changing a flexible polynucleotide segment into a more rigid one
  may for instance, strongly enhance the bending fluctuations in another
  domain. 
\end{itemize}

\section{Materials and Methods}
\label{Sec:Materials-Methods}

\subsection{Synthesis of oligonucleotides}

The dsDNA 60bp HPLC purified oligonucleotides GAGE6, GAGE6\_1, GAGE6\_2, and
GAGE6\_3 (Fig. 1) were produced by Integrated DNA Technologies and resuspended
in 150 mM KCl, 20 mM HEPES (pH 7.4), 1 mM DTT, 5\% (v/v) glycerol, and 5 mM
MgCl$_2$ buffer to a DNA concentration of 70-80 {$\mu$}M (exact concentrations
specified in Table 1).

\subsection{Size Exclusion Chromatography-Small Angle X-ray Scattering
  (SEC-SAXS)}

SAXS data for all oligonucleotides were collected on the SAXS/WAXS beamline at
the Australian Synchrotron using an inline SEC-SAXS sheath-flow setup
\cite{KIRBY,RYAN}.
Data were all collected by loading 50 $\mu$L of
DNA at 70-80 $\mu$M in a buffer of 150 mM KCl, 20 mM HEPES (pH 7.4), 1 mM DTT,
5\% v/v glycerol, and 5 mM MgCl$_2$.
All samples were analysed on a pre-equilibrated Superdex 200 Increase 5/150
column (GE Healthcare) with UV absorbance at 260 nm and 280 nm monitored
alongside scattering. Data reduction to $I(q)$ vs $q$ scattering profiles was
carried out using SCATTERBRAIN (software for acquiring, processing and viewing
SAXS/WAXS data at the Australian Synchrotron) correcting for sample
transmission and solvent scattering and placed on an absolute intensity scale
using a water standard. Subsequent data processing and analysis were performed
using the ATSAS suite \cite{PETOUKHOV}. As discussed by Trewhella et
al. \cite{TREWHELLA} SCATTERBRAIN outputs the uncertainty of intensity
measurements as 
2$\sigma$. For subsequent analysis, these uncertainties were transformed to
$\sigma$ for all 
data sets such that all metrics used for comparing models and experimental
data had conventional interpretations.
Supplementary figure S4 shows that this does not affect the
shape of the distance distribution functions $P_{\mathrm{exp}}(r)$ which
  this method uses for analysis (see sections 2.3 and 2.4).
For all SEC-SAXS data, regions with
self-consistent, non-nucleic acid frames, prior to the elution of the main
peak were averaged and taken as solvent scattering with the program CHROMIXS
\cite{CHROMIXS}. The sample
scattering was then taken as the average of frames with similar radius of
gyration $R_g$ values within the peak corresponding to DNA.
(Supplementary figure S1)
To additionally assess the monodispersity of the frames across each
chromatography peak, the molecular weight of the individual frames was
calculated in BioXTAS Raw \cite{HOPKINS2017}. This was done using the Vc
method \cite{RAMBO2013} with the molecule type set to 'RNA' rather
than protein and the average window size set to 1 (Supplementary figure S2).
Guinier analysis was
performed using ATSAS 4.0 \cite{MANALASTAS}. Distance
distribution analysis was performed using ATSAS 3.2.1 \cite{MANALASTAS}.
To assist in the selection of the optimisation parameters for our
various $P(r)$  functions the program AutoGNOM which is
part of the ATSAS 
package  for the analysis of SAXS data \cite{PETOUKHOV} was also used.
To assess the role of regularisation in the stability of our main
features in $P_{\mathrm{exp}}(r)$, we have performed a
cross-comparison between the optimal
solutions derived from AutoGNOM-4 from ATSAS 2.7.2 and AutoGNOM-5 from ATSAS
3.2.1.

\begin{figure*}
  \begin{center}
    \includegraphics[width=17cm,clip]{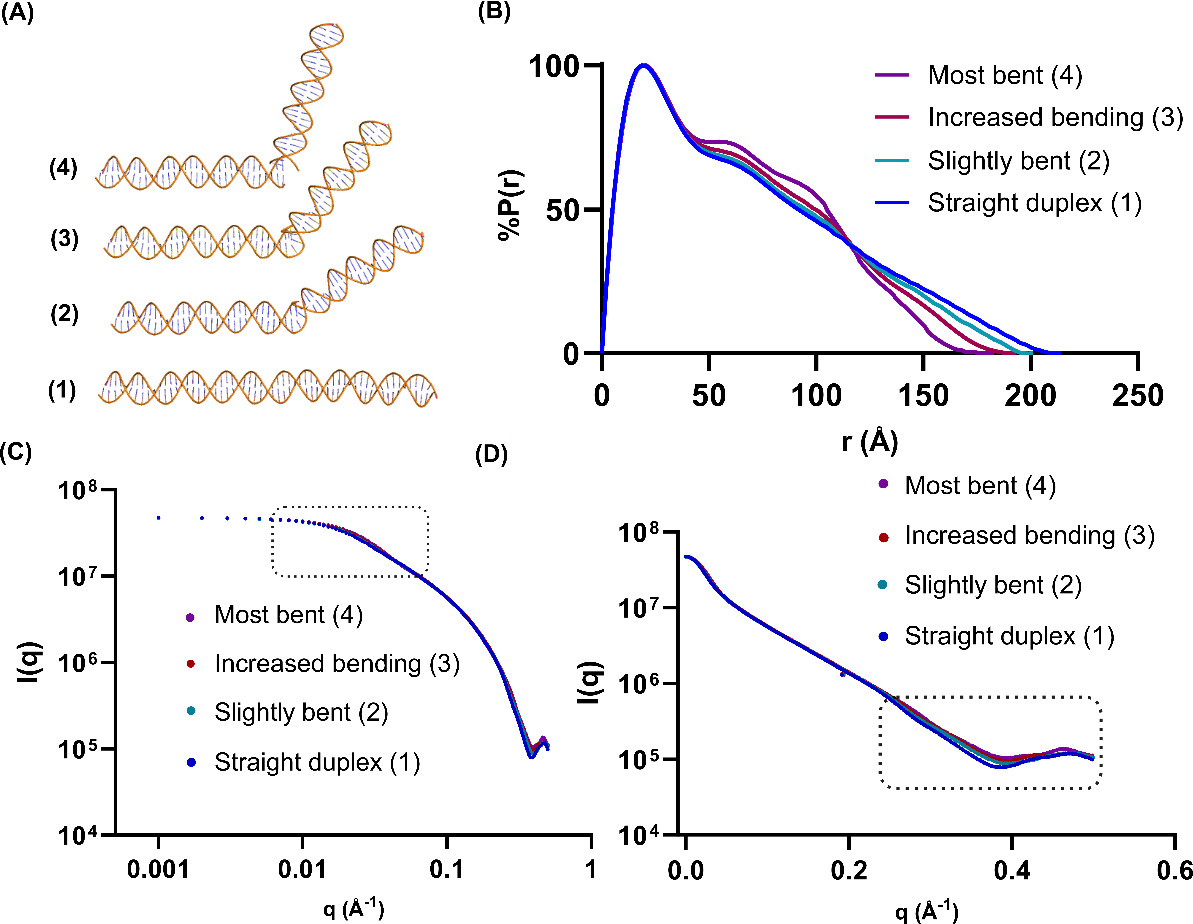}
  \end{center}
    {Figure 2: Illustration of the effect of DNA bending on the shape of
$P(r)$.  (A) A group of simple 60 bp DNA models, the first of which was
created by Alphafold 3 \cite{ABRAMSOL}.  Models 2-4 were created by variably
bending the first alphafold model manually at base-pair 34. 
(B) Pair-distance probability distribution functions $P(r)$ for these 4
conformations computed with CRYSOL \cite{SVERGUN995}. ( C/D) the simulated
scattering from the 4 duplexes is shown as $\log(I)$ vs. $\log(q)$ (C) or
$\log(I)$ vs. $q$ (D). The areas where the computed scattering patterns differ
across both types of plots are indicated by the dotted box.}
  \label{fig:figdeux}
\end{figure*}

\subsection{Choosing real space instead of reciprocal space for  analysis}

Scattering experiments measure the intensity as a function of the scattering
vector $\mathbf{q}$.
Coherent elastic scattering is determined by the correlations in the
local scattering density of the sample and expressed by the Fourier transform
of the autocorrelation function of the scattering density. Moreover, for
dilute molecular solutions, where the orientation and position of individual
molecules are uncorrelated, the only coherent contribution to the scattering
comes from the internal structure of the molecules. As the scattering
represents the ensemble and time-averaged scattering of all molecules
illuminated by the beam, all orientational information is lost and the
scattering patterns are radially symmetric. Thus, scattering data from dilute
monodispersed solutions depends only on the modulus $q$
of the scattering vector.
The Fourier transform in polar coordinates reduces to a one-dimensional
integral \cite{Sivia}
\begin{equation}
  \label{eq:iofq}
  \nonumber
  I(q) \propto \int_0^{D_\mathrm{max}} P(r) \frac{\sin qr}{qr} dr
\end{equation}
where $P(r)$ is the pair-distribution function of the scattering particle.
This function can be related to the probability of finding two scattering
points separated by a distance between $r$ and $r + dr$. The shape of the
$P(r)$ function is characteristic of the shape of the molecule and
depends on the input parameter $D_{\mathrm{max}}$, which is the maximum
distance between two scattering points within a given molecule.  Whilst
$D_{\mathrm{max}}$ is selected and optimised during the fitting process, it
also reflects a real physical property, which is the maximum linear dimension
of the molecule. Data are collected in reciprocal space $q$, while we are
interested in 
conformations i.e. properties related to the geometry of the molecules in real
space $r$. The pair-distance distribution function $P(r)$ can be derived from
$I(q)$ by inverse Fourier transform
\begin{equation}
  \label{eq:getpr}
  \nonumber
  P(r) \propto r^2 \int_{0}^{\infty} q^2 I(q) \frac{\sin qr}{qr} dq
\end{equation}
meaning that the data in real space $r$ and reciprocal space $q$ are related
by a 
mathematical transform and are therefore formally equivalent. However, it is
evident from Eq. (\ref{eq:iofq})
that knowledge of the $P(r)$ function permits the
calculation of $I(q)$, however, from Eq. (\ref{eq:getpr})
it is clear that the inverse is not
true. Computation of $P(r)$ from $I(q)$ involves an infinite integral, yet the
scattering data is measured over a finite $q$-range, thus, a direct
calculation of inverse
Fourier transformation of $I(q)$ will be dependent on the $q$-range over which
the 
data is measured. To address this issue, an indirect Fourier transformation of
the scattering data is used, whereby a set of basis functions are combined to
generate a $P(r)$ function, from which $I(q)$ can be calculated and compared to
the experimental scattering data. The coefficients of the basis functions are
then optimised (subject to various restraints and constraints) against the
scattering data to yield the $P(r)$ function for the measured $I(q)$ data. The
real space representation of the scattering data is more intuitive to most
people, and importantly, subtle differences in reciprocal space
$I(q)$ can often be more obvious in real space.
  To this point, Fig. 2 plots the theoretical $P(r)$ and
$I(q)$ profiles of a short segment of DNA made of two straight double helices
connected by a variety of local bends. In the theoretical data the structural
differences due to bending lead (Figure 2A) to an accumulation of differences
at specific $r$ values and a notable qualitative change in the shape of $P(r)$
(Figure 2B). Whilst the theoretical scattering for these different bent
structures is only subtly different below $q = 0.23\,$\AA$^{-1}$
(Figure 2C, D) the
scattering begins to show discernible differences above this threshold in
the theoretical profiles (Figure 2D). However, when considering real
experimental data (see Figure 5 for an example) above this threshold, the
differences may become increasingly difficult to discern due to rising noise
levels as the data extends along q. In contrast to the smooth theoretical
curves, the experimental scattering profile also becomes much more diffuse
at higher q, possibly reducing sensitivity to structural differences among
such conformations when relying on this part of the scattering data.
{\em This example illustrates a crucial choice for the successful analysis
  of 
  our SAXS data:} to detect the structural properties of the DNA molecules it
is much more efficient and meaningful to fit $P(r)$ in place of $I(q)$.
This can be easily understood from the general relationship between real space
and reciprocal space. In this study, we are looking for DNA features which
could affect the binding of proteins, such as curvature or flexibility on the
scale of the binding domain. These features are {\em localised}
structures in real
space, corresponding to {\em extended} structures in the Fourier-transformed
reciprocal space, where they are spread over a broad q range, and therefore
are harder to detect.

\medskip
The radius of gyration can be calculated directly from the $P(r)$ and can be
generally determined more accurately when compared to the Guinier
approximation \cite{GLATTER1977} as a significantly larger region of reciprocal
space is used to calculate the $P(r)$ \cite{BLANCHET}.  

\medskip
To perform a comparison of the measured datasets in reciprocal space the data
were re-gridded in PRIMUS with a spacing of $\Delta q = 0.001\;$\AA$^{-1}$.
The re-gridding
was carried out using a common $q$-range between all four experiments of
$0.0068-0.4918\;$\AA$^{-1}$.
The ‘data compare’ function in PRIMUS was then used to
assess the statistical similarity of the datasets in reciprocal space.
To focus on the $q$-range that gives rise to the features
that we have observed in
$P_{\mathrm{exp}}(r)$, the maximum value of $q$ of the data used for the test
was set to $0.18\;$\AA$^{-1}$.
It computes a $p$-value by checking for patterns in the cross-correlation
matrix between datasets, which does not depend on the values of the
experimental errors \cite{FRANKE2015}. While $p > 0.95$ means that two
datasets are almost identical, low values of $p$, typically below a threshold
of $0.01-0.05$, indicate that the datasets are statistically independent. Such
values 
are generally used to identify outliers, such as sample misloading,
denaturing or radiation damage \cite{MANALASTAS}.
Intermediate values of $p$ indicate some correlations between the
datasets. Supplementary Table 1 shows that the $p$-values obtained by
comparing the GAGE6 data successively with data recorded for GAGE6\_1,
GAGE6\_2, GAGE6\_3 are 0.493, 0.280, 0.078. This indicates
decreasing correlations
between the data sets, which are consistent with the nature of our
samples. They derive from the same GAGE6 sequence, with only local sequence
changes, growing from GAGE6\_1 to GAGE6\_3. Therefore one should not expect
that such samples lead to completely statistically independent datasets as
it occurs for outliers due to experimental problems. These $p$ values are
interesting because our analysis (see Results section 3) detects a large-bending
domain in GAGE6\_1 and GAGE6\_2 in a portion
of the sequence which is not far from
the bent domain of detected for GAGE6, which is compatible with a fairly large
$p$-value whereas this strongly bent domain disappears for
GAGE6\_3, which should decrease the correlation with the GAGE6 dataset leading
to a lower $p$-value. {\em The statistical analysis appears therefore to
  provide a very good preview of the results presented in Sec.~3.}

\subsection{Derivation of $P_{\mathrm{exp}}(r)$ from experimental data}

The derivation of the pair-distance distribution function (Supplementary
figure S3) 
$P_{\mathrm{exp}}(r)$ from $I(q)$ versus $q$ via an indirect Fourier transform is
performed with the program GNOM of the ATSAS package \cite{MANALASTAS}.
The program requires several input parameters which can variably affect the
results. Some of these relate to the experimental setup, and others have to be
selected. Two of these selected parameters are crucial to the transform
process: the maximum diameter of the sample $D_{\mathrm{max}}$ ,
and the regularisation parameter $\alpha$.
\smallskip

The parameter $D_{\mathrm{max}}$ can be estimated to a good accuracy from the
properties of the samples.  According to the typical base-pair distance of
$3.3 - 3.4$ \AA\ in the B-form of DNA \cite{SAENGER}, a fully rigid ideal
60bp duplex should have a maximum dimension $D_{\mathrm{max}}$ of $\approx$
205 \AA\ (excluding its hydration shell). Using this as an estimate for what
our approximate maximal dimension should be in solution, one of the criteria
was that the selection of $D_{\mathrm{max}}$ for each function falls within
10\% (184.5 to 225.5 \AA) of this value, as there can often be an inherent
uncertainty associated with $D_{\mathrm{max}}$
which can be difficult to quantify \cite{TREWHELLA}. As
shown by the supplementary figure S5, varying $D_{\mathrm{max}}$ over a range
of $20\;$\AA\ (which is very significant in comparison with the length of the
DNA sequences) only has a minute effect on the shape of $P_{\mathrm{exp}}(r)$.
Furthermore, as small errors in $D_{\mathrm{max}}$ can alter $P(r)$
slightly \cite{GRANT2015}
persistence of features in the respective functions in the
face of varying $D_{\mathrm{max}}$ serves as an additional indicator of the
robustness of the solution.

\smallskip
The choice of the regularisation parameter $\alpha$ is more delicate. The
program GNOM includes a process called AutoGNOM which optimises a Total Quality
Estimate (TQE) that combines an ensemble of different criteria
\cite{SVERGUN1992} such
as the discrepancy between the reconstructed Fourier space data and the
experimental data, the degree of oscillations in the calculated
$P_{\mathrm{exp}}(r)$, and the stability of the solution versus the variation of
$\alpha$. A convergence process optimises the TQE. Throughout the development of
the ATSAS package the method has evolved and currently, the
more contemporary version of GNOM-5 introduced a new smoothness
criterion for $P_{\mathrm{exp}}(r)$
that was not part of GNOM-4 but is now used in addition to the other metrics
to calculate the TQE. The best
definition of the TQE and the best convergence procedure may depend on the
type of sample. It is likely that a sample with a well defined shape such as a
folded protein may have different requirements from a long flexible polymer
like DNA, as these often are molecules that may have very different
conformations in solution. 
Thus, in spite of the long development period of ATSAS, the
selection  of the optimal $\alpha$ parameter remains a challenge. This is why
we have compared the results provided by both GNOM-5 and GNOM-4. They are
plotted in Fig. S6 of the supplementary material which also shows how the
TQE of GNOM-4 depends on $\alpha$ for all the samples.

For GAGE6 the results of GNOM-5 and GNOM-4 are very close to each other for
the shape of $P_{\mathrm{exp}}(r)$ and its estimated error bars. For GAGE6\_1
and GAGE6\_3 both optimisation procedures converge towards similar shapes
although some differences are noticeable. For GAGE6\_2, GNOM-5 converges to
$\alpha = 0.01084$, i.e. a very weak regularisation leading to
$P_{\mathrm{exp}}(r)$ showing significant oscillations and large error
bars. GNOM-4 converges to $\alpha = 1.31$, i.e. a fairly strong regularisation
with a very smooth $P_{\mathrm{exp}}(r)$ and small error bars. Figure S6
shows that, for GAGE6\_2, the TQE has a wide plateau where it stays almost
constant when $\alpha$ varies. This explains why the determination of the
optimal $\alpha$ is particularly difficult for this sample. Therefore, for
some samples, the calculation of $P_{\mathrm{exp}}(r)$ raises some doubts 
concerning the validity of the results. However, Supplementary
figures S6 and S7 show that $\alpha$ essentially
determines the {\em magnitude} of the humps of $P_{\mathrm{exp}}(r)$ but has
little effect on their {\em position} with respect to $r$.
This is important because Fig. 2
shows that the amplitude of a local DNA bend affects the magnitude of the
humps in $P_{\mathrm{exp}}(r)$  but not their position in $r$, which is
determined by the location of the bend along the sequence.

This suggests that a possible indetermination of $\alpha$ could have little
influence of the main DNA feature that we are trying to detect in our
analysis, i.e. how the sequence of DNA locally modifies its propensity for
bending, which may affect protein binding. As shown later, even
for the case of
sequence GAGE6\_2, this conjecture is confirmed by the calculation
presented in Sec. 3.3, although
it is the sequence for which the evaluation of $\alpha$ and the shape of
$P_{\mathrm{exp}}(r)$ is the hardest. Supplementary figure S10 shows that the
same is true for all sequences.

\medskip
This shows that SAXS data for flexible polymers, which can hardly be
distinguished from each other in reciprocal space, even with mathematical
tools, can contain detailed information which is revealed when the results
are expressed in real space. The above factors together with the simulated
$P(r)$ functions which show bending a dsDNA model will cause bumps to appear
in the function, the fact that AT tracts have been proven to cause bending
of dsDNA, the good fits of each $P(r)$ function to the data (Supplementary
Figure S8) and their reasonable TQE (Total Quality Estimate) scores strongly
support the validity of the information collected in real space. This is
further confirmed by the correlation between the sequence of our DNA samples
and the local properties that we detect in our analysis based on
$P_{\mathrm{exp}}(r)$.

\medskip
Simulated $P(r$) functions (Figure 2) were generated using CRYSOL (Svergun et
al., 1995) on the first Alphafold 3 model and the 3 subsequent manually bent
models, applying fake 1\% errors to the theoretical
intensity and proceeding with the standard
derivation of $P(r)$ in Primus.  

\subsection{Analysis: from $P_{\mathrm{exp}}(r)$ to the conformation of the
  samples.}

\subsubsection{General concept}

The pair distribution function, which describes the probabilities of all the
atom-atom distances in the conformational ensemble of a sample, summarises a
large amount of information. Even in the absence of conformational
heterogeneity,  ab initio structural modelling against high-quality data from
monodisperse solutions does not yield unambiguous structures. Add the
complication of conformational heterogeneity and extraction of structural and
conformational information from scattering data is not possible in the absence
of complementary information from other biophysical techniques. Indeed, there
are numerous packages available for modelling ensembles of flexible molecules
using SAXS data, but all rely on having high-resolution structural information
of all components in the flexible system \cite{BERNARDO,SCHNEIDMAN,TRIA}.
Similarly to flexible protein solutions \cite{MARTIN2021},
DNA in solution is an ensemble of
different structures that are evolving over time, and beyond a few helical
turns, a DNA molecule undergoes large-scale fluctuations. Our analysis
proceeds along similar lines, taking advantage of the knowledge of the general
properties of DNA to build a suitable ensemble of conformers. Our method was
first developed to search for kinked-DNA conformations in SAXS and SANS data
on a nucleosome positioning sequence \cite{MPKINK}. However, it has
been significantly improved for the present study with the set of data
obtained on a series of related sequences with additional steps now added to
the method. A polymer model of DNA is used to widely scan the conformational
space of the sample. Ensembles of conformations which have a pair-distance
distribution probability sufficiently close to $P_{\mathrm{exp}}(r)$
are selected and
saved. Then a subsequent analysis selects a subset of optimal conformations
which are then statistically analysed to extract the main features of the DNA
sample such as its local distribution of bending or twist angles and their
standard deviations. In some cases, small differences in the sequence are
found to lead to large changes in sample properties, which might be relevant
for protein binding affinity or the selection of a binding site.
Although a
random scan of conformations is, in principle, possible provided it is
sufficiently broad, it would be highly inefficient. The choice of a suitable
polymer model that reflects the properties of DNA is important. However the
model must not bias the results, and therefore it does not include any
assumption on the local properties of the sample.

\smallskip
The approach used by Peterlin \cite{PETERLIN} to extract the persistence
length of DNA from light scattering data gives some hints on possible
approaches for SAXS data. Peterlin introduced a simple polymer model, the
Kratky-Porod model, which consists of a chain of rigid segments connected by
flexible joints with an energy that depends on the angle $\theta$
between the two
joined segments. For this simple model, the theoretical calculation of the
scattering function is possible within some limits and its comparison with the
experimental data determines the appropriate model parameter, from which the
persistence length can be derived. The same idea was refined by Schellmann
\cite{SCHELLMAN1974}
for a better description of DNA flexibility which took into
account dihedral angles $\varphi$
between the planes defined by pairs of consecutive
segments. However, to allow for analytical calculations, he only considered
two limiting cases, either $\varphi$ is constrained to zero,
i.e.\ the model evolves in a plane, or $\varphi$ is totally random. 
For DNA the reality lies between these two limits which correspond to either
an infinite torsional rigidity ($\varphi = 0$)
or zero torsional rigidity ($\varphi$ random,
evolving freely without any energy cost). This is why we have chosen an
extended Kratky-Porod model with an added energy contribution for the dihedral
angles in order to describe not only the persistence length of DNA but also
its proper torsional rigidity.

\setcounter{figure}{2}
\begin{figure}
  \null \vspace{0.5cm}
  \centering
    \includegraphics[width=7.5cm]{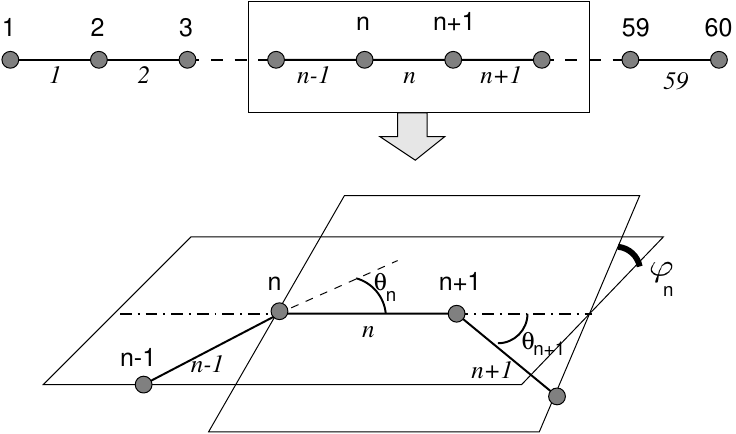}
  \caption{
  The DNA polymer model used in the analysis of the SAXS data.}
  \label{fig:model}
\end{figure}

\subsubsection{The DNA polymer model}

The DNA polymer model is shown in Fig.~3. {\em It is not
concerned with the internal structure of the DNA helix but only with the
conformation of the line going through the center of the double helix.}
It consists of $N$ objects representing the
base pairs, each separated by the base pair distance in DNA $a=3.34\;$\AA. Its
$N-1$ segments can be viewed as a generalised Kratky-Porod model. Its
conformation is defined by the bending angles $\theta_n$, $n=2, \ldots ,N-1$
at sites $n$, between segment $n-1$ and $n$ ($ -\pi < \theta_n \leq \pi$)
and the dihedral angles $\varphi_n$,
$n=2, \ldots ,N-2$, between the planes defined by segments $[n-1,n]$ and
segments $[n,n+1]$ ($ -\pi/2 < \varphi_n \leq \pi/2$). The reference state for
this model is a straight DNA helix, with 11 base pairs per turn, i.e.\ an
angle of about $33^{\circ}$ between them. The $\theta$ and $\varphi$ angles
describe the deviations with respect to this structure.

The thermal fluctuations of the backbone are controlled by the variation of its
bending and torsional energies $E_{\theta}$ and $E_{\varphi}$
\begin{equation}
  \label{eq:hamil}
  H = E_{\theta} + E_{\varphi} =
  \sum_{n=2}^{N-1} K (1 - \cos \theta_n) + \sum_{n=2}^{N-2} C (1 - \cos
  \varphi_n) \; .
\end{equation}
The constants $K$ and $C$ set the scale of the bending and torsional
energies. They are assumed to be the same along the full model, and 
the bending and twist energies are minimal
for $\theta=0$ and $\varphi=0$,
i.e.\ {\em the model is homogeneous, with no permanent bending or torsion, in
  order 
to avoid any bias towards a specific conformation other that a straight,
untwisted, DNA molecule.}

\smallskip
The numerical calculations use dimensionless variables. Lengths are measured
in units of $a = 3.34\,$\AA\
the DNA base pair distance so that the dimensionless length of
the segments is unity. The energy is measured in units of $k_B
T_{\mathrm{room}}$ where $k_B$ is the Boltzmann constant and
$T_{\mathrm{room}}$ is room temperature in Kelvin chosen as $298\;$K.  The two
model parameters $K$ and $C$ are chosen to correspond to the physical
properties of DNA.

\smallskip
The constant $K$ determines the persistence length of the
model, i.e.\ the length $L_p$
over which unit vectors $\mathbf{R_1}$, $\mathbf{R_2}$ tangent to the helix
axis lose collinearity according to $\langle \mathbf{R_1} \cdot \mathbf{R_2}
\rangle = \langle \cos \theta \rangle \approx \exp(-l_{12}/L_p)$
where $\langle \rangle$ designates a statistical average and $l_{12}$ is the
distance along the molecule between the two contact points of $\mathbf{R_1}$
and $\mathbf{R_2}$.
It has been determined by a broad variety of methods, from
macroscopic evaluations \cite{PETERLIN}, to single-molecule experiments
\cite{SMITH-SB92} and, although it depends on external conditions \cite{LU-Y}
the consensus for standard conditions is $L_p=500\;$\AA, i.e.\ in the unit
length of our model $L_P = 500/3.34 \approx 150$ units. For the Kratky-Porod
model, in the limit of $L_p$ much larger than the length of a segment, $L_p$
is given by $L_p = a K / k_B T$, so that, with our dimensionless variables
where $a$ and $k_B T$ are unity (for experiments at room temperature), we get
$L_p = K$. Therefore the choice $K=150$ gives a model with a persistence
length appropriate for DNA.

\smallskip
The constant $C$ determines the torsional rigidity $\xi$
of the model in the linear
limit, i.e.\ for small torsional angles $\varphi$ for which $E_{\varphi} \approx
(1/2) C \varphi^2$ and a torque $\tau = \partial E_{\varphi}/\partial \varphi = C
\varphi$.
In this limit a DNA molecule of length $L$ behaves as an elastic rod whose
torsional torque grows with torsion as $\tau = (\xi/L) \varphi$. For DNA
measurements of $\xi$ vary from about $200\;\mathrm{pN \cdot nm^2}$ to
$480\;\mathrm{pN \cdot nm^2}$.
Topoisomer distributions of small circles, yielding $\xi \approx
300\;\mathrm{pN \cdot nm^2}$ are generally considered as the most reliable
measurements \cite{BRYANT}. In the context of our model, for a base pair,
i.e.\ $L=a=3.34\;$\AA, we get $C = \xi/a = 898\, \mathrm{pN \cdot nm}$.
When this
value, homogeneous to energy (force $\times$ length), is expressed in our
energy unit $k_B T_{\mathrm{room}}$ we get the
dimensionless value $C \approx 218$. Taking into account
the broad error bars in the experiments, 
we have used the value $C=200$ in our calculations as a realistic value to
describe the torsional modulus of DNA.

\medskip
Conformations of this model at room temperature are generated using
Monte-Carlo simulations. The first segment $\mathbf{R_1}$ is chosen
arbitrarily 
along the $z$ axis of our reference frame, and a second segment $\mathbf{R_2}$
is added in
the $x,z$ plane by selecting a random value of $\theta_2$ in the range
$-\pi < \theta_2 \leq \pi$.
From two vectors we can generate a local frame $x',y',z'$, using
$\mathbf{R_2}$ as the $z'$ axis, choosing an orthogonal $x'$ axis in the
$\mathbf{R_1}$,$\mathbf{R_2}$ plane and completing by a third axis $y'$ to get
an orthonormal frame. A new segment is added in this local frame by
selecting 
$\theta$ in the range $-\pi < \theta \leq \pi$ and the dihedral
angle $\varphi$ in the
range $-\pi/2 < \varphi \leq \pi/2$,
and so on until the $59$ segments corresponding to our
$60$-base-pair DNA samples have been added. Each time a new segment has been
defined the contribution $E_{\theta} + E_{\varphi}$ that it adds to the energy is
compared to $k_B T$ ($k_B T = 1$ in our dimensionless units) and
accepted or rejected with the Metropolis algorithm \cite{METROPOLIS1953}.
Once a conformation of $59$ segments has been completed the probability
distribution  $P(r)$ of the pair distances between its 60 beads
is obtained by a calculation which directly derives from its definition.
We compute all pair distances $r_{ij}$, $i=1,\ldots,N$, $j=1,\ldots,N$,
$i \not= j$, i.e.\ $N(N-1)/2 = 1770$ values. Then $P(r)$ is derived from
a normalised histogram of these distances by counting how many
of them lie within each of $N+1$ boxes
of size $a$ centred on the values $r_k = (k - \frac{1}{2})a$
(we use real distances,
not the dimensionless distances for comparison with experimental data)
$ P(r_k) = {n_k}/{\sum n_k}$,
where $n_k$ is the number of  $r_{ij}$ values that fall within box $k$. 
The least-squares distance $S$ between $P(r)$ and $P_{\mathrm{exp}}(r)$,
the normalised distribution function obtained from experimental data,
interpolated to get its values at points $r_k$, is given by
$S = \big[ \sum_k \big( P(r_k) - P_{\mathrm{exp}}(r_k) \big)^2 \big]^{1/2}$.

In
this sum only the values $r_k > 25\;$\AA\ are considered to take into account
the limitations of the DNA model, restricted to the backbone, which cannot
describe DNA at very small distances.
The experimental pair-distance probability distribution
function $P_{\mathrm{exp}}(r)$ contains many short-distance contributions
which are not relevant for the overall conformation of the molecule. These
include the distances between atoms belonging to the same base-pair for which
$r \le d \approx 25\;$\AA, where $d$ is the diameter of the molecule,
including bound water. Other short distances bind atoms belonging to
neighbouring or next-neighbouring base pairs, which are also within $r
\lesssim 25\;$\AA. The model, which represents one base-pair by a single point
does not include all those contributions, and even slightly above this
threshold it tends to underestimate $P_{\mathrm{exp}}(r)$. On the other hand,
for larger distances, the backbone gives a meaningful picture of
the conformation of a DNA molecule and its flexibility, which is the aim of
our analysis. If $S$ is smaller than a
predefined threshold $S_0$ the configuration is saved in a file for further
analysis. To make sure that we only keep the conformations
that provide an optimal
fit to our experimental data, $S_0$ is selected so that only $10$ to $50$
conformations are saved 
for one million generated conformers. To ensure the thorough exploration of
the conformational space of samples a calculation can generate up to
$5\,10^8$ trial conformations, which is possible even with modest computing
facilities because the model is sufficiently simple, with only two parameters
per DNA base pair. The same analysis with an all-atom DNA model would be
beyond even the best computing facilities.

\subsubsection{Analysis of the saved conformations}

Among the saved conformations, we select the 1000 which have the smallest
least square difference $S$ between $P(r)$ and $P_{\mathrm{exp}}(r)$ and
we use these ``best'' conformations for statistical analysis.
An individual
conformation has little physical meaning because at room temperature a
60~bp DNA molecule is sufficiently flexible to allow its shape to fluctuate
widely. Nevertheless its properties are reflected in the statistics of these
fluctuations.

In the context of protein binding, DNA curvature and flexibility are important,
therefore the first relevant set of statistics concern the bending angles
$\theta_n(i)$ where $n$ is the index of base pairs along the sequence and $i$
($i = 1, \ldots 1000$) denotes the conformations.

To demonstrate these principles we foreshadow in
Figure 4 as an example of the average over the 1000 best
conformations of the absolute values $|\theta_n(i)|$ of the local bending
angles, which measure the deviations from a straight double helix, i.e.\
\begin{equation}
  \label{eq:avgtheta}
  \overline{|\theta_n(i)|} = \frac{1}{1000} \sum_{i=1}^{1000} | \theta_n(i) |   
\end{equation}
as well as their standard deviations
\begin{equation}
  \label{eq:stddev}
  \sigma_{\theta} = \left[
    \frac{1}{1000} \sum_{i=1}^{1000} \left( | \theta_n(i) |
      - \overline{|\theta_n(i)|} \right)^2 \right]^{1/2}
  \end{equation}
We also show the average of $|\theta_n(i)|$ restricted to bending angles which
exceed a threshold $\theta_0 =   5^{\circ}$ to stress the domains which are
prone to large bending. The average diehdral angle $\overline{|\varphi_n(i)|}$
and the standard deviation $ \sigma_{\varphi}$ are computed with equations
similar to Eqs.~(\ref{eq:avgtheta}) and (\ref{eq:stddev}).

Figure 4 shows a typical pattern from
SAXS results: it is symmetrical with respect to the centre of the sequence.
This was expected because the structure factor
and the probability distribution
function $P_{\mathrm{exp}}(r)$ result from an average of all the possible
orientations of the molecules in space. Analysing $P_{\mathrm{exp}}(r)$ in
terms of conformations has already reduced this three-dimensional degeneracy
and brought the data along a line corresponding to the DNA chain.
However, SAXS data
alone do not allow us to identify a particular end of the sequence.  To
discuss the results we would like to compare the sequence with its base pairs
identified by an index $p$, $p=1 \ldots 60$, and the
angles along the model $| \theta_n|$, $n=1, \ldots 60$,
but we face a dilemma that the experiment cannot solve:
it cannot tell us whether $p=1$ corresponds to $n=1$ or whether it corresponds
to $n=60$. The two peaks may be real, or one of them may be an artefact
because the statistical analysis mixes sequences with $n=p$ and others with $n
= 60 - p +1$ so that we sum up data with a single peak on the right and data
with a single peak on the left.

\begin{table*}
  \parbox{\linewidth}{Table 1: Sample information and characterization - Small
    Angle X-Ray Scattering parameters} 
\label{tab:saxsparam}
\begin{center}
\small
  \begin{tabular}{|p{0.24\linewidth}|%
    p{0.16\linewidth}|p{0.16\linewidth}|p{0.16\linewidth}|p{0.16\linewidth}|}
\cline{1-5}
\textbf{(A) Sample details} & \textbf{GAGE6} & \textbf{GAGE6\_1} &
\textbf{GAGE6\_2} & \textbf{GAGE6\_3}\\
\cline{1-5}
Organism & Homo Sapiens & Variant of original GAGE6 sequence &
Variant of original GAGE6 sequence & Variant of original GAGE6 sequence \\
\cline{1-5}
Source (Catalogue No. or \par reference) & doi:10.1021 \par
                                           /acs.biochem.2c00192. & 
 & &  \\
\cline{1-5}
\textbf{Scattering particle composition} &  &  &  & \\
  \cline{1-5}
DNA/RNA(s) & 60 bp dsDNA & 60 bp dsDNA & 60 bp dsDNA & 60 bp dsDNA \\
Stoichiometry of components & Single component & Single component &
Single component & Single component \\
\cline{1-5}
\textbf{Sample environment\ /configuration} & & & & \\
 \cline{1-5}
 Solvent composition & 150 mM Kcl, 20 mM HEPES (pH 7.4), 1 mM DTT,
 5\% glycerol and 5 mM MgCl$_2$ &
150 mM Kcl, 20 mM HEPES (pH 7.4), 1 mM DTT,
 5\% glycerol and 5 mM MgCl$_2$ &
150 mM Kcl, 20 mM HEPES (pH 7.4), 1 mM DTT,
 5\% glycerol and 5 mM MgCl$_2$ &
150 mM Kcl, 20 mM HEPES (pH 7.4), 1 mM DTT,
 5\% glycerol and 5 mM MgCl$_2$ \\
\cline{1-5}
Sample temperature ($\strut^{\circ}$C) & $25^{\circ}$C &
$25^{\circ}$C & $25^{\circ}$C & $25^{\circ}$C \\
\cline{1-5}
In beam sample cell & Co-flow & Co-flow & Co-flow & Co-flow \\
\cline{1-5} 
\textbf{Size Exclusion Chromatograpy SEC-SAXS} & & & & \\
    \cline{1-5}
    Sample injection concentration ($\mu$M) & 69.7 & 70 & 75 & 81 \\
    \cline{1-5}
    Sample injection volume, mL & 0.05 & 0.05 & 0.05 & 0.05 \\
    \cline{1-5}
    SEC column type & S200 5/150 & S200 5/150 & S200 5/150 & S200 5/150 \\ 
    \cline{1-5}
    SEC flow rate, mL/min & 0.4 & 0.4 & 0.4 & 0.4 \\
    \cline{1-5}
  \end{tabular}

  \begin{tabular}{|p{0.24\linewidth}|%
    p{0.16\linewidth}|p{0.16\linewidth}|p{0.16\linewidth}|p{0.16\linewidth}|}
\cline{1-5}
\textbf{(B) SAXS Data Collection} & \textbf{GAGE6} & \textbf{GAGE6\_1} &
\textbf{GAGE6\_2} & \textbf{GAGE6\_3}\\
\cline{1-5}  
Data acquisition/reduction software & Scatterbrain 2.82 &
Scatterbrain 2.82 & Scatterbrain 2.82 & Scatterbrain 2.82 \\
\cline{1-5} 
Source/instrument description or reference & 
Australian Synchrotron SAXS/WAXS beamline &
Australian Synchrotron SAXS/WAXS beamline &
Australian Synchrotron SAXS/WAXS beamline &
Australian Synchrotron SAXS/WAXS beamline \\
\cline{1-5}
Wavelength (nm) & 0.10781 & 0.10781 & 0.10781 & 0.10781 \\
\cline{1-5}
Camera length (mm) & 3000 & 2385 & 2385 & 2385 \\
\cline{1-5}
Measured $q$-range ($q_{min} - q_{max}) \,$\AA$^{-1}$ &
$0.0045-0.49$ & $0.0068-0.62$ & $0.0068-0.62$ & $0.0068-0.62$ \\ 
\cline{1-5}
Method for scaling intensities & Absolute scaling against water &
Absolute scaling against water & Absolute scaling against water &
Absolute scaling against water \\
\cline{1-5}
Exposure time(s), number of exposures. &
frames $183-194$ selected (12 1 second frames) &
frames $156-162$ selected (7 1 second frames) &
frames $163-164$ selected (2 1 second frames) &
frames $166-172$ selected (7 1 second frames) \\
\cline{1-5}
\end{tabular}

\begin{tabular}{|p{0.24\linewidth}|%
    p{0.16\linewidth}|p{0.16\linewidth}|p{0.16\linewidth}|p{0.16\linewidth}|}
\cline{1-5}
  \textbf{(C) SAXS-derived structural parameters}
  & \textbf{GAGE6} & \textbf{GAGE6\_1} &
\textbf{GAGE6\_2} & \textbf{GAGE6\_3}\\
\cline{1-5}
Methods/Software & ATSAS 4.0 & ATSAS 4.0 & ATSAS 4.0 & ATSAS 4.0 \\
\cline{1-5}
\textbf{Guinier Analysis} & & & & \\
\cline{1-5}
$I(0) \pm s$ (cm$^{-1}$ a.u) & $0.005 \pm 0.00013$ &
$0.015 \pm 0.00026$ & $0.013 \pm 0.00028$ & $0.016 \pm 0.00018$ \\
  \cline{1-5}
$R_g \pm s \,$(\AA) & $49.89 \pm 3.55$ &
$60.22 \pm 2.23 $ & $54.21  \pm 2.55 $ & $56.37 \pm 1.52 $ \\  
\cline{1-5}
$\mathrm{min} \le q R_g \le \mathrm{max}$ limit (or data point range) &
$0.37 - 0.96 $ & $0.52 - 1.01 $ & $0.42 - 1.06 $ & $0.38 - 1 $ \\
\cline{1-5}
Linear fit assessment (fidelity in PRIMUS) & 1 & 0.71 & 0.9 & 0.97 \\
\cline{1-5}
Point range & $5 - 22$ & $5 - 22$ & $5 - 22$ & $5 - 22$ \\ 
\cline{1-5}
$\mathbf{PDDF/P(r)}$ \textbf{analysis} & ATSAS 3.2.1 &
ATSAS 3.2.1 & ATSAS 3.2.1 & ATSAS 3.2.1 \\
\cline{1-5}
$I(0) \pm s$ (cm$^{-1}$ a.u) & $ 0.00509 \pm 0.0001516 $ &
$ 0.01459 \pm 0.0001741 $ & $ 0.01250 \pm 0.0002808 $ & 
$ 0.01554 \pm 0.0002117 $ \\
\cline{1-5}
$R_g \pm s \,$(\AA) & $ 55.18 \pm 1.829 $ & $ 54.24 \pm 1.031 $ &
$ 55.37 \pm 1.860 $ & $ 55.90 \pm 1.380 $ \\
\cline{1-5}
$d_{\mathrm{max}} \,$(\AA) & 195 & 192 & 184.7 & 215 \\
\cline{1-5}
$q$-range (\AA$^{-1}$) & $ 0.0080 - 0.1704 $ & $ 0.0104 - 0.1554 $ &
$ 0.0077 - 0.1518 $ & $ 0.0068 - 0.1536 $ \\
\cline{1-5}
$P(r)$ fit assessment (total quality estimate) & 0.67 (reasonable) &
 0.70 (reasonable) & 0.67 (reasonable) &   0.68 (reasonable) \\
\cline{1-5}
Alpha & 1.308 & 0.552 & 0.413 & 1.48 \\
\cline{1-5}
\end{tabular}

\end{center}
\end{table*}

\begin{table*}
  \parbox{\linewidth}{Table 1 continued:
    Small Angle X-ray scattering data collection
  parameters} 
\label{tab:saxsparam2}
\begin{center}
\small
\begin{tabular}{|p{0.24\linewidth}|%
    p{0.16\linewidth}|p{0.16\linewidth}|p{0.16\linewidth}|p{0.16\linewidth}|}
\cline{1-5}
  \textbf{(D) Scattering particle size}
  & \textbf{GAGE6} & \textbf{GAGE6\_1} &
\textbf{GAGE6\_2} & \textbf{GAGE6\_3}\\
\cline{1-5}
Methods/Software & ATSAS 3.2.1 & ATSAS 3.2.1 & ATSAS 3.2.1 & ATSAS 3.2.1 \\
\cline{1-5}
\textbf{Volume estimates} & & & & \\
  \cline{1-5}
Volume (\AA$^{-3}$) (MoW method) & 43073 & 69629 & 69523 & 79329 \\  
\cline{1-5}
\textbf{Molecular weight (M) estimates (kDa)} & & & & \\
\cline{1-5}
From chemical composition & 37.101 & 37.105 & 37.111 & 37.116 \\
\cline{1-5}
From SAXS concentration independent method (volume of correlation method) &
30.06 & 31.61 & 35.53 & 34.95 \\
  \cline{1-5}
  \end{tabular}

\begin{tabular}{|p{0.24\linewidth}|%
    p{0.16\linewidth}|p{0.16\linewidth}|p{0.16\linewidth}|p{0.16\linewidth}|}
\cline{1-5}
  \textbf{(D) Scattering particle size}
  & \textbf{GAGE6} & \textbf{GAGE6\_1} &
\textbf{GAGE6\_2} & \textbf{GAGE6\_3}\\
\cline{1-5}
\textbf{(E) Data deposition} & & & & \\
  \cline{1-5}
SASBDB ID: & \textbf{SASDXH7} & \textbf{SASDV87} & 
\textbf{SASDV97} & \textbf{SASDVA7} \\   
 \cline{1-5}
\end{tabular}
  \end{center}
\end{table*}

{\em However, if the sequence has some intrinsic asymmetry
 this gives an additional piece of information
  which can be combined with the SAXS data to solve
the dilemma} by examining the conformations one by one and reversing
some of them
so that the peak of large $|\theta_n|$ is always on the same side, for
instance to the right that we call ``forward'' conformation. The algorithm to
achieve what we call the ``orientation'' of the conformations is the following.
We identify
the indices $n_1$, $n_2$ on both sides of the large $n$ peak, for instance
$36$ and $48$ in the case of the GAGE6 sample of Fig.~4.
The mirror positions are $n'_1 = N - n_2 +1$ and $n'_2 = N - n_1 +1$
i.e. $n'_1 = 13$ and $n'_2=25$ for the example above.
Then we scan again the 1000 best conformations and compute for each of them
$s = \sum |\theta_n|, n=n_1 \ldots n_2$ and $s' = \sum |\theta_n|, n=n'_1 \ldots
n'_2$. If the difference between $s$ and $s'$ is
significant, which is evaluated by calculating $\Delta s/s = |s'-s|/s$, 
then one of the two peaks is ficticious.
If $s > s'$ this conformation is already in the forward
orientation. If $s' > s$ the conformation has to be reversed by switching
indices $n$ into $n' = N - n + 1$.  After this check, constructing a plot
similar to Fig.~4 with all conformations oriented
``forward'', only one of the two peaks remains. As shown in the Results section,
this is what happens for the 4 samples that we studied. When reversing a
conformation, we also reverse the indices for the dihedral angles $\varphi_n$
paying attention to the fact that, while $n$ varies from $2$ to $N-1$ for
$\theta_n$, it varies from $2$ to $N-2$ for $\varphi_n$ so that, instead of $n'$
one has to use $n'' = N - n$ for $\varphi$. Besides the elimination of spurious
signals, the orientation process improves the quality of the averaging because
it becomes real averaging over the 1000 best configurations instead of the
superposition of two sets of data averaged over about 500 conformations.

\begin{figure}
  \null \vspace{0.5cm}  
  \centering
    \includegraphics[width=7.5cm]{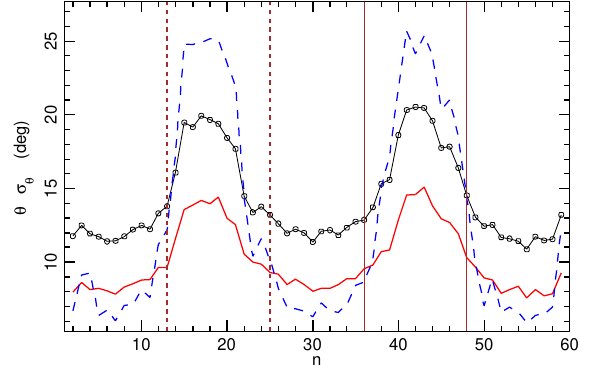}
  \caption{
    Statistics of the local bending angles $|\theta_n(i)|$ over the 1000
    best saved conformations for the GAGE6 sample: the red full line shows
    $\overline{|\theta_n(i)|}$. The black line with circles shows the same
    average, limited to the bending angles which are above $\theta_0 =
    5^{\circ}$, and the dashed blue curve shows the standard deviation of
    $|\theta_n(i)|$.
    The full brown vertical lines show the positions $n_1=36$, $n_2=48$ which
    have been used for the 'orientation' of the conformations and the dashed
    brown vertical lines show the mirror positions $n'_1=13$ and $n'_2=25$. 
    }
  \label{fig:nonorient}
\end{figure}

After this process has been completed we know that all the
conformations generated by the model have the same orientation with respect to
the sequence, i.e.\ that for all of them $n=1$ corresponds to $p=1$ or all of
them correspond to $p=60$. It does not yet tell us which of the two choices is
correct, and further data processing cannot answer that. However, now that we
have a reliable statistical analysis we can compare $\overline{|\theta_n(i)|}$
with the sequence. We can expect that the domains with large bending angles
are more likely to be found in regions which are richer in AT base pairs,
usually more flexible than the GC-rich regions. For instance for
sequence GAGE6 (Section 3.2), after
orientation we detect a single peak around sites $38 - 44$
(Fig.~6). In the sequence the domain $p = 38 - 44$ is a large
continuous domain with only AT pairs. This suggests that, for this sequence,
the correct choice is $n=p$, i.e. the case for which $n=1$ corresponds to
$p=1$.

\medskip
This ``orientation'' step is important to improve the quality of the
statistical analysis,
but it is also essential to discuss the results, when we start to examine how
the local properties of the sample compare with the local sequence. It is a
step that should be included in all the SAXS studies on polymer chains. In
theory, if the chains
don't have any intrinsic asymmetry, they could be
tagged by an additional short domain at 
one end, chosen for its properties to be easily identified in the SAXS data
and so allow for their orientation.

\medskip
The analysis can then be extended to detect other properties of the sample. In
the Results section, we show how the dihedral angles vary along the sequence,
and we also calculate the probability $P(n,|\theta|)$ showing, for each site,
the distribution of the bending angles among the 1000 best conformations, and
similarly for $P(n,|\varphi|)$. These figures are very helpful to get a view
of the likely conformations of the samples in solution and to try to
understand why some sequences may be favourable for protein binding. Because
long polymers, like DNA, fluctuate extensively in solution, {\em a statistical
analysis of their most likely conformations is crucial to understanding
biological phenomena. It provides information that usefully complements
high-resolution structures of protein-DNA complexes, which often only show a
snapshot of the final conformational state after binding.}

\section{Results and Discussion}

\subsection{Small Angle X-ray Scattering data}

The chromatograms computed with the program CHROMIXS \cite{CHROMIXS}
indicated the presence of one single main peak
eluting in a similar position for all the duplexes, with some small shoulders
on either side of the peak (Supplementary figure S1). All eluted peaks had
260:280nm UV absorbance ratios consistent with that of pure dsDNA
(Supplementary figure S1). Each dsDNA peak had some instability in predicted
$R_g$ values across the eluted peak, possibly because of neighbouring
peaks. However, frames were selected for analysis from regions where the
$R_g$
was stable across consecutive frames within approximately
$5\,$\AA\ or less (Supplementary figure S1). In the case of GAGE6, the $R_g$
across the chosen frames varied from
$45.3$ to $50.78\,$\AA\
with an average of $48.08\,$\AA. For GAGE6\_1 the $R_g$ varied from
$49.9\,$\AA\ to $52.9\,$\AA\  with an average of $52.0\,$\AA.
For GAGE6\_2 the $R_g$ of the chosen
frames ranged from $53.9-54.0\,$\AA\
with the $R_g$ in the surrounding frames varying
from $51.5- 54.0\,$\AA.
In the case of GAGE6\_3, the variation in the predicted $R_g$
across the chosen frames was minimal, ranging from
$49.3-50.6\,$\AA\  with an average of $50.1\,$\AA. 

\smallskip
The calculation of the molecular weight of the frames across the
chromatography peak worked well for GAGE6\_1, GAGE6\_2, and GAGE6\_3; however,
it was unsuccessful for GAGE6, perhaps due to the fact that this experiment
is slightly noisier than the others. As shown in Supplementary figure S2,
the molecular weight of the frames across the peaks are in agreement with
the expected molecular weight of a dsDNA duplex of this size ($\approx
37$kDa). This 
demonstrates that at least in the case of 3 of the 4 chromatograms, the
samples are clearly monodisperse. Despite this calculation not working for
the GAGE6 sample, the $R_g$ values of the chosen frames for the GAGE6
experiment are in close agreement, demonstrating that this sample is also
likely to be monodisperse.

\medskip
All scattering shown as $\log_{10} (I)$ vs. $\log_{10} (q)$
plots demonstrated good quality
data with a plateau in the Guinier region (Fig. 5). When constrained to
$q.Rg \lesssim 1$
(suitable for elongated molecules), the Guinier fits had high fidelity
scores and acceptable distributions of residuals. The fits also provided
$R_g$ 
values similar to the expected value for a straight 60bp dsDNA duplex as
simulated with CRYSOL ($R_g = 56.10\,$\AA) (Fig. 5). The distance distribution
functions of the various oligonucleotides from this study highlight some
robust features across each variant. Depending on the duplex the shape of the
$P_{\mathrm{exp}}(r)$ function changes,
with characteristic bumps in the GAGE6, GAGE6\_1, and
GAGE6\_2 functions which then almost
disappear in GAGE6\_3 as shown in Figs. 6, 7, 8,
9 and 10  in the next sections. The $D_{\mathrm{max}}$ values
for all functions were within the range of ±10\% of the theoretical
$D_{\mathrm{max}}$ of a
60 bp duplex. Supplementary figure S5 shows that deviation of
$D_{\mathrm{max}}$ from the
optimal value given by our procedure only leads to small variations in
$P_{\mathrm{exp}}(r)$.
Additionally, Supplementary Figure S6 shows that despite variation of
the GNOM regularisation parameter the bumps in our respective
$P_{\mathrm{exp}}(r)$ functions
persist. The $R_g$  values derived from the individual $P_{\mathrm{exp}}(r)$
functions were more or
less consistent with those from the Guinier approximation. We noted minor
inconsistencies between $P_{\mathrm{exp}}(r)$ and Guinier-derived $R_g$
for GAGE6 and GAGE6\_1. The
details concerning SAXS parameters are summarised in Table 1.

\begin{figure*}
  \begin{center}
    \includegraphics[width=15.5cm]{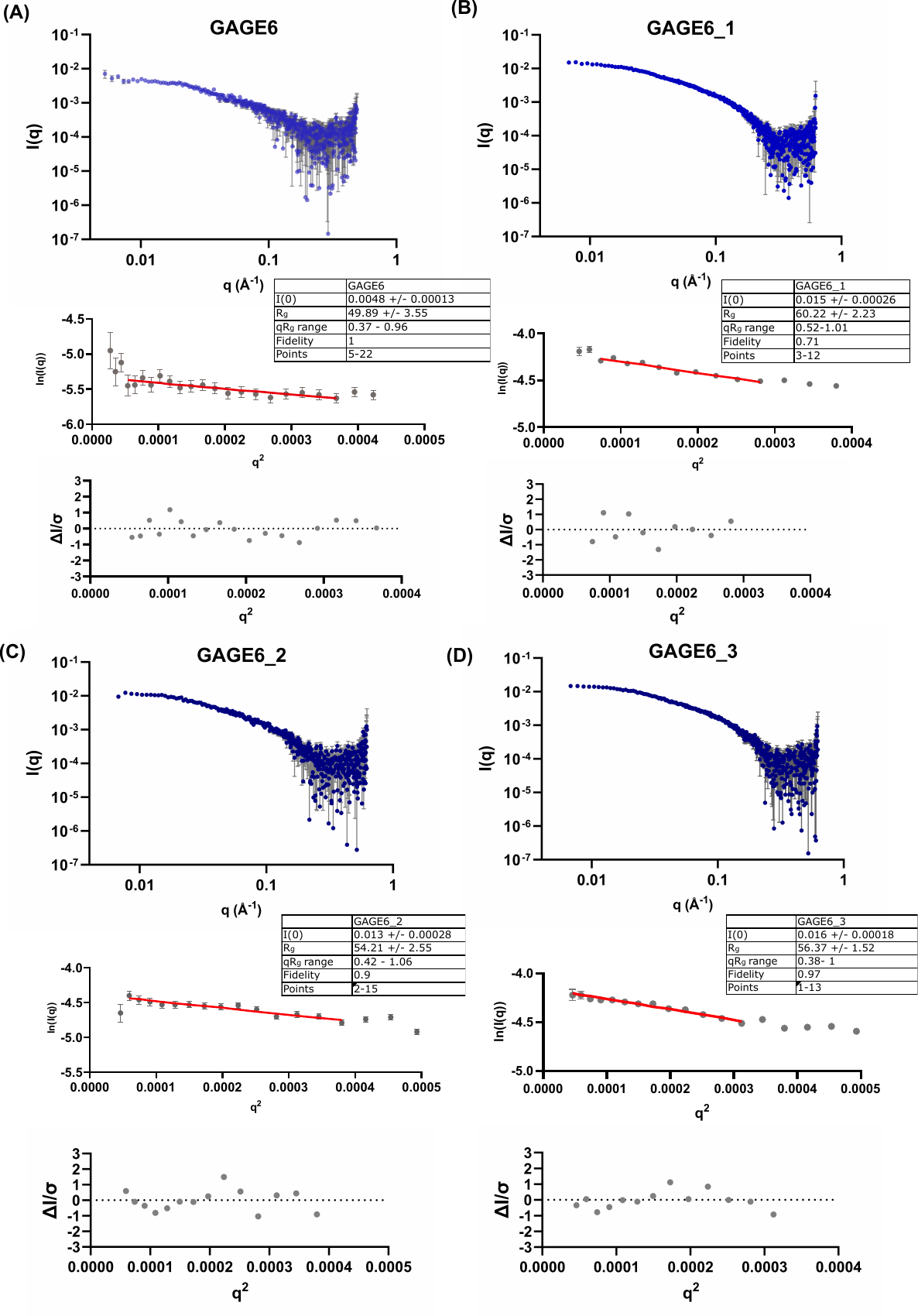}
  \end{center}

  \parbox{16cm}%
  {Figure 5:
    Scattering data and Guinier fits of the various GAGE6 oligonucleotides:
    $I(q)$ versus $q$ plot, Guinier plot ($\ln[I(q)]$ vs $q^2$), and Guinier
    residual plots for (A) GAGE6, (B) GAGE6\_1, (C) GAGE6\_2, (D) GAGE6\_3.
    Guinier fit parameters and results are tabulated on the figure.
  }
  
  \label{fig:scattering}
\end{figure*}

\subsection{Sequence GAGE6}

Let us start with the GAGE6 sequence, which is the leading sequence of the
series since it is a native promoter sequence that is bound by SFPQ. 
Figure 6A
compares the experimental pair-distance distribution function
$P_{\mathrm{exp}}(r)$ with $P(r)$ calculated from the 1000 best conformations
saved after running Monte-Carlo simulations to generate conformations with the
polymer model. The agreement between the two distributions is evaluated by
computing
\begin{equation}
  \label{eq:chi2}
  \chi^2 = \frac{1}{N_r - 1} \sum_1^{N_r} \left[ \frac{P(r_j)
      - P_{\mathrm{exp}}(r_j)}{\sigma_j} \right]^2 \; ,
\end{equation}
where $r_j$ are the points where $P_{\mathrm{exp}}(r)$ has been calculated,
$\sigma_j$ the experimental error at these points (estimated
by the GNOM program),
and $N_r$ the number of calculation points. The summation is restricted to $r_j
> 25\;$\AA\ because
the polymer model cannot describe the internal structure
of DNA, which dominates $P_{\mathrm{exp}}(r)$ for $r < 25\;$\AA\ 
as discussed above (section 2.5.2). The curve deduced
from the selected conformations of the polymer model (Figure 6A)
provides a very good
description of the experimental curve in the whole domain where the polymer
model is valid, indicating in this instance that the polymer model ensemble
can very accurately reproduce $P_{\mathrm{exp}}(r)$.

\begin{figure*}
  \begin{center}
    \includegraphics[width=16cm]{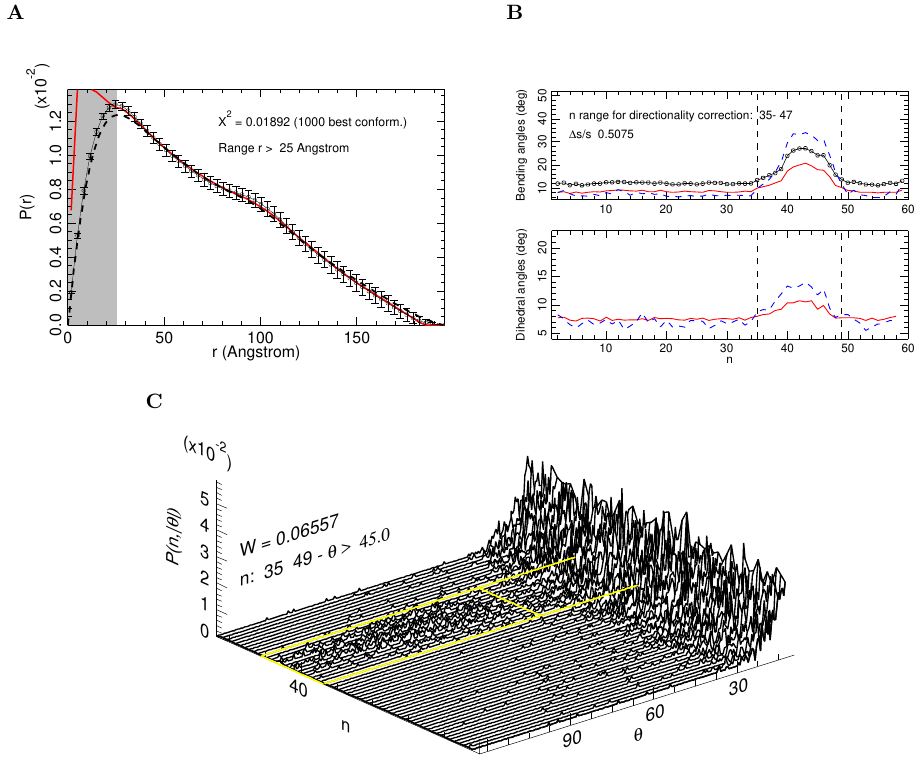}
  \end{center}
  \parbox{17cm}{Figure 6:
    Sequence GAGE6: (A) Comparison between
    $P_{\mathrm{exp}}(r)$ calculated with AutoGNOM-5
    (small circle points and thin black
curve with error bars) and $P(r)$ calculated for the 1000 best conformations
generated theoretically (thick red curve). The grey region, for $r <
25\,$\AA\  is
the low-$r$ domain in which the model cannot describe the internal structure of
the double helix. The thick black dashed line shows the function
$P_{\mathrm{exp}}(r)$ obtained with AutoGNOM-4 (see section 2.4)
(B) Sequence GAGE6:
Upper part: Statistics of the absolute value of the local bending angles
$|\theta_n(i)|$ over the 1000 best-saved conformations for the GAGE6 sample,
after orientation of the conformations as discussed in the Materials and
Methods section. The red full line shows the average over $i$ of
$|\theta_n(i)|$ versus $n$.  The black line with circles shows the same
average, limited to the bending angles which are above $|\theta| = 5^{\circ}$,
and the dashed blue curve shows the standard deviation of $|\theta_n(i)|$.
Lower part: Statistics of the absolute values of the dihedral angles The red
full line shows the average over $i$ of $|\varphi_n(i)|$ versus $n$ and the
the dashed blue curve shows their standard deviation.
(C) Sequence GAGE6: Three-dimensional plot of the
normalised probability distribution function $P(n,|\theta|)$.
The data are not shown
for small bending angles, for which the probabilities are the highest, to
better show the data at larger $\theta$.
The value $W$ marked on the figure shows the
fraction of bending angles $|\theta_n|$ above the value shown
below $W$ within the domain indicated by the two values of $n$. 
  }
  \label{fig:resu-gage6}
\end{figure*}

\medskip
Figure 6B shows the statistics of the bending angles
$|\theta_n|$ and dihedral angles $|\varphi_n|$ with their averages over
conformations and theit standard deviations.
For the bending angles this figure is
analogous to Fig.~4 except that the conformations have been
oriented as described in the Materials and Methods section. Only one of the
double maxima that were
forming a mirror image with respect to the middle of the
sequence has subsisted, which confirms that the symmetric pattern was created
through symmetry inherent in the SAXS measurement. The value $\Delta s / s$
marked on the figure corresponds to $(s' - s)/s$, where $s$ and $s'$ were the
local sums of the bending angles in the two mirror domains used to orient the
conformations. Their differences were significant, suggesting that one of the
two mirror images was fictitious, which is confirmed by the figure as one of
them has been fully erased in the orientation step.

The three-dimensional plot of Fig.~6C shows the probability $P(n, |\theta|)$
of the bending angles $|\theta_n|$ for each site $n$ along the
sequence. These distributions are normalised
  \begin{equation}
    \label{eq:normhtheta}
    \int_0^{\pi} P(n, |\theta|) \; \mathrm{d}|\theta| = 1
    \quad {\mathrm{ for~each~}}n
  \end{equation}
  which allows us to calculate the fraction $W$ of the bending angles
  $|\theta_n|$  above a some value in a given domain.
  The result $W = 0.06557$ is
shown in Fig~6C for $ 35 \le n \le 49$ and a lower value of
$\theta$ of $45^{\circ}$. Thus, although most of the bending angles are well
below $30^{\circ}$, in some regions of the sequence there are about $6\%$ of
the  bending angles above $45^{\circ}$.
The statistics clearly show that one region of the sequence is special, with a
much higher flexibility and larger bending than elsewhere.
The colour view of the three-dimensional probability $P(n, |\theta|)$ (Fig.~6C),
(see Supplementary material Fig.~S9) 
suggests that the large bending angles for $ 35 \le n
\le 49$ are separated from the bulk of smaller bending angles, as if they
result from some permanent bend in this region and not simply from large
amplitude thermal fluctuations. As this part of the probability
distribution only concerns
$6\%$ of the bending angles, the resolution of the analysis of the SAXS data
does not however allow us to make a definitive statement on this point.

\medskip
We would like to stress that these statistical results which provide precise
information on the local properties of the sample, are extracted from a
probability distribution function $P_{\mathrm{exp}}(r)$
which looks rather featureless although it
results from a large ensemble of pair distances. It is remarkable that this
smooth looking, almost straight function, actually encodes so much
information. Now that we have oriented the conformations we can also
compare the statistical results with the actual sequence of the sample. There
is still an unknown point: should the sequence listed in
Fig.~1 be oriented from 1 to 60 or from 60 to 1 ?

\medskip
The GAGE6 sequence is asymmetric, with large AT-rich domains on the 3’ end,
while the 5’ end is more GC-rich. The AT base pairs, linked by only two
hydrogen bonds, are generally more flexible than the GC pairs linked by three
hydrogen bonds because their weaker bonding allows greater freedom in their
local conformations. This effect is also enhanced by easier fluctuational
openings, even at room temperature \cite{DNAFLEX}.
Therefore it seems reasonable to assume that the $n = 60$ side of our
diagrams corresponds to the 3’ end. With this assignment, we can notice that
the domain $n = 35 \ldots 47$ that we used for the orientation step is almost
completely made of AT pairs with only two, separate, GC pairs inside
it. Therefore it is not surprising to find that this domain is highly flexible
as found in the analysis of the SAXS data. Thus the results of our analysis of
the SAXS data suggest that this polymer modelling pipeline is able to detect
specific features at the scale of a few base pairs. Its validity is
strengthened by the results on the fluctuations of the dihedral angles, which
show a maximum in their standard deviations which coincides with the maximum
of the bending angles, although it is slightly broader. In the linear chain of
joined segments that we use to analyse the data there is no geometrical
constraint which links bending and twist. But in DNA there is such a
constraint due to the double helical structure. It is easy to check with a
mechanical device such as two wires twisted together to make a helix that a
large amount of bending tends to locally untwist the device, simply by
geometrical effects. {\em The fit of our polymer model to
$P_{\mathrm{exp}}(r)$ detects this
phenomenon although it is not built into our hypothesis.}

\medskip
However, although the sequence strongly contributes to determining the local
flexibility of the sample the results also point to subtle effects associated
with some cooperativity along the double helix. In the GAGE6 sequence, sites
$38-44$ make a large continuous domain with only AT pairs and therefore it is
not surprising to find higher bending angles in this region as pointed out
above. But the GAGE6 sequence also has a 4-AT-pairs domain at sites $28-31$
which are not within the region where we detect  larger bending
angles. Similarly, the 5-AT-pairs domain at sites $50-54$ also show fairly low
bending angles.

\medskip
The sequence gives a hint on the local flexibility of DNA but the analysis of
the SAXS data suggests that experiments which are sensitive to the
conformation of DNA tell us more. This may have many biological
implications. Within cells and viruses, DNA has to be tightly bent or kinked
at many sites to allow for compact packaging of large amounts of genetic
material \cite{WIDOM1984}. Many studies have tried to address the question of
whether DNA bending arises from metastable bent structures or more flexible
“joints” in dsDNA. Previously solved structures of dsDNA AT-rich dodecamers
\cite{STEFL2004,HIZVER2001} suggest that bent duplexes are
stable and are maintained in a bent conformation by several molecular
features. This notion was supported by a fluorescence polarisation experiment
\cite{CHIRICO2001} that also argued for a permanent static bend in some
AT-rich dsDNA sequences. A review by Harteis and Schneider
\cite{HARTEIS2014} on the structure of DNA summarised that whilst AT tracts can
introduce an intrinsic bend, they in fact somewhat increase the rigidity of
such sequences due to certain stabilising effects that accommodate the bending
of DNA: propellor twisting of bases, narrow minor groove, wider major groove
\cite{STEFL2004,HIZVER2001} as well as non-Watson-Crick hydrogen
bonds forming down the major groove. Rather interestingly, the bending of
AT-rich DNA was also shown to be sensitive to the order of the bases, with
A4T4 vs. T4A4 steps having different effects, the former causing a global bend
in the helix, and the latter forming a straighter helix \cite{STEFL2004}.
DNA has been considered to act more flexibly when a “kink” defined as
the local unstacking of base pairs causing a change in the orientation of the
helix appears \cite{HARTEIS2014}. Pyrmidine-purine (TA; CA) steps
reportedly exhibit the lowest base-stacking energy and can behave more like
flexible hinges that can cause DNA bending.

\medskip
In addition to these structural effects, further studies have also examined the
fluctuational opening of DNA base pairs or the formation of “bubbles” and
whether these could be a source of flexibility in DNA. Gueron et al.
\cite{GUERON1987}
described the probability of the fluctuational opening of base
pairs as low via imino-proton exchange. More recently Theodorakoupolous and
Peyrard \cite{DNAFLEX} reviewed this concept and theorised
that below the melting point of a duplex, an increase in temperature towards
physiological temperature may increase the opening of base pairs enough to act
as flexible hinges in dsDNA, which may have wider implications for DNA
structure and recognition by proteins.

\begin{figure*}
  \begin{center}
    \includegraphics[width=16cm]{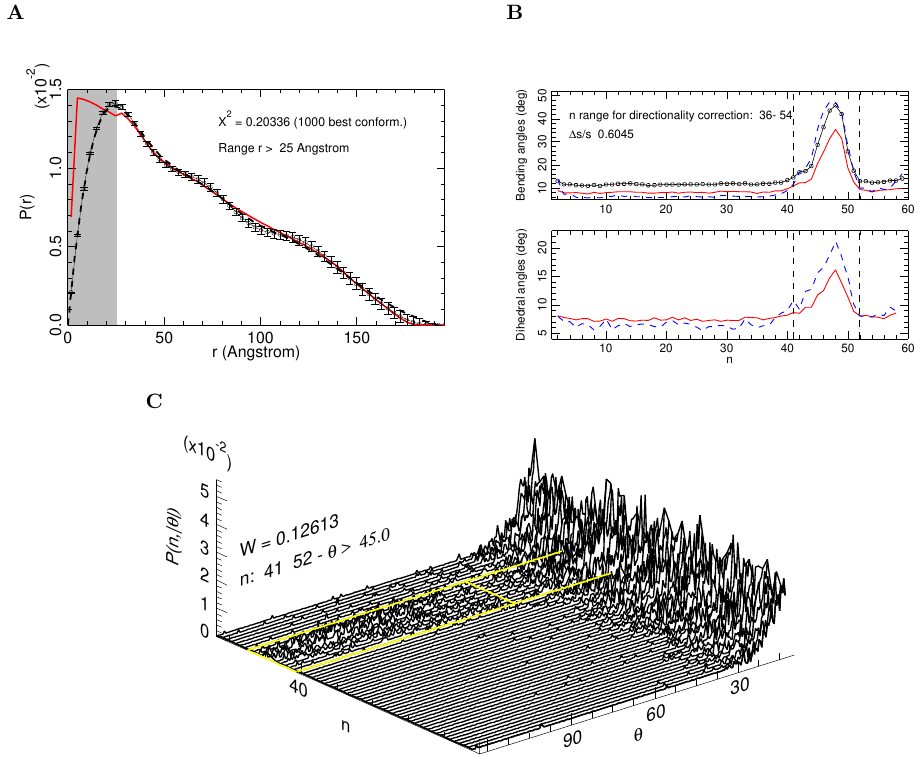}
  \end{center}
  \parbox{17cm}{Figure 7:
    Sequence GAGE6\_1: (A) Comparison between
    $P_{\mathrm{exp}}(r)$ calculated with AutoGNOM-5
    (small circle points and thin black
curve with error bars) and $P(r)$ calculated for the 1000 best conformations
generated theoretically (thick red curve). The grey region, for $r <
25\,$\AA\  is
the low-$r$ domain in which the model cannot describe the internal structure of
the double helix. The thick black dashed line shows the function
$P_{\mathrm{exp}}(r)$ obtained with AutoGNOM-4 (see section 2.4)
(B) Sequence GAGE6\_1:
Upper part: Statistics of the absolute value of the local bending angles
$|\theta_n(i)|$ over the 1000 best-saved conformations for the GAGE6 sample,
after orientation of the conformations as discussed in the Materials and
Methods section. The red full line shows the average over $i$ of
$|\theta_n(i)|$ versus $n$.  The black line with circles shows the same
average, limited to the bending angles which are above $|\theta| = 5^{\circ}$,
and the dashed blue curve shows the standard deviation of $|\theta_n(i)|$.
Lower part: Statistics of the absolute values of the dihedral angles The red
full line shows the average over $i$ of $|\varphi_n(i)|$ versus $n$ and the
the dashed blue curve shows their standard deviation.
(C) Sequence GAGE6\_1: Three-dimensional plot of the
normalised probability distribution function $P(n,|\theta|)$.
The data are not shown
for small bending angles, for which the probabilities are the highest, to
better show the data at larger $\theta$.
The value $W$ marked on the figure shows the
fraction of bending angles $|\theta_n|$ above the value shown
below $W$ within the domain indicated by the two values of $n$. 
  }
  \label{fig:resu-gage61}
\end{figure*}

\medskip
Alongside the fluctuational opening of the duplex, further unwinding of DNA
regions is also important. Taking a simple physical model of 2 interwound
strands and bending it sufficiently causes unwinding of the 2 strands at the
point of bending. Recently, Chandrashekhar et al. \cite{CHANDRASEKAR2024}
demonstrated via a minicircle assay that DNA can locally unwind in response to
bending. This has implications for gene regulatory and DNA packaging elements,
as the bending of DNA by any of the aforementioned mechanisms may also
contribute to the unwinding of the duplex. 

In a SAXS experiment, it can be complicated to delineate between interrelated
effects such as stable, rigid/bent polyAT structures, flexibly kinked DNA as a
result of unstacked base pairs, the unwinding of strands, or the fluctuational
opening of base pairs. This is particularly true given that SAXS is relatively
speaking a low-resolution structural technique. However, having shown that we
can quantitatively detect the bending of DNA in solution we hypothesised that
such events could be described by the complementary study of a series of
related sequences derived from GAGE6.

\subsection{Sequences GAGE6\_1, GAGE6\_2, GAGE6\_3}

This series of three sequences was derived from GAGE6 by incrementally
removing its longest AT-rich domains to replace them with GC-rich
segments. This allowed us to test two things
\begin{itemize}
\item[(i)] to check how the analysis of the SAXS data detects these local
  changes and how they influence the conformations of the samples, and  

\item[(ii)] to try and understand the properties of the GAGE6 sequence
  that allow it to bind SFPQ. The first sequence of the series, GAGE6\_1, only
  differs from GAGE6 by four bases in sites $28-31$. The TTTT sequence in GAGE6
  has been replaced by GCGC in GAGE6\_1. 

\end{itemize}

Figure 7 shows the equivalent data for the GAGE6\_1 variant to Figure 6 for
GAGE6. Figure 7A shows that, although the change in the sequence is very
localised, the shape of $P_{\mathrm{exp}}(r)$
is significantly modified: the single bump around
$100\,$\AA\ in GAGE6 splits into two lumps around $70$ and
$125\,$\AA\  for GAGE6\_1. Our
analysis shows that this difference is due to a large change in the
distribution of $\theta$ angles (Fig. 7B). The broad domain
with large $\theta$ angles
covering sites $36 - 46$ in GAGE6 is now narrower, and restricted to sites
$44 - 52$ 
only, but is also much sharper as the maximum of $\theta_n(i)$
reaches $36^{\circ}$ instead of
about $20^{\circ}$ for GAGE6.
The dihedral angles show the same behaviour, with large
twist fluctuations being restricted to the same narrow domain.

\medskip
Figure 7C confirms the presence of large bending angles in a narrow domain and
the integration of $P(n,|\theta|)$ for $41 \le n \le 52$ and $\theta >
45^{\circ}$ shows that the fraction 
$W$ of bending angles $|\theta_n|$
above $45^{\circ}$ has drastically increased from 6.4\% to
12.6\%. The GAGE6\_1 sequence therefore has a high probability of being sharply
bent in a narrow region. It is interesting to notice the qualitative analogy
between the experimental pair-distance distribution function for GAGE6\_1 and
the double-bump simulated $P(r)$ (Fig. 2) obtained with an atomistic model of
sharply bent DNA.  This strengthens the notion that $P(r)$ has a high
sensitivity to molecular conformations, particularly bending, with sharp
bending generating a function with two bumps.

\begin{figure*}
  \begin{center}
    \includegraphics[width=16cm]{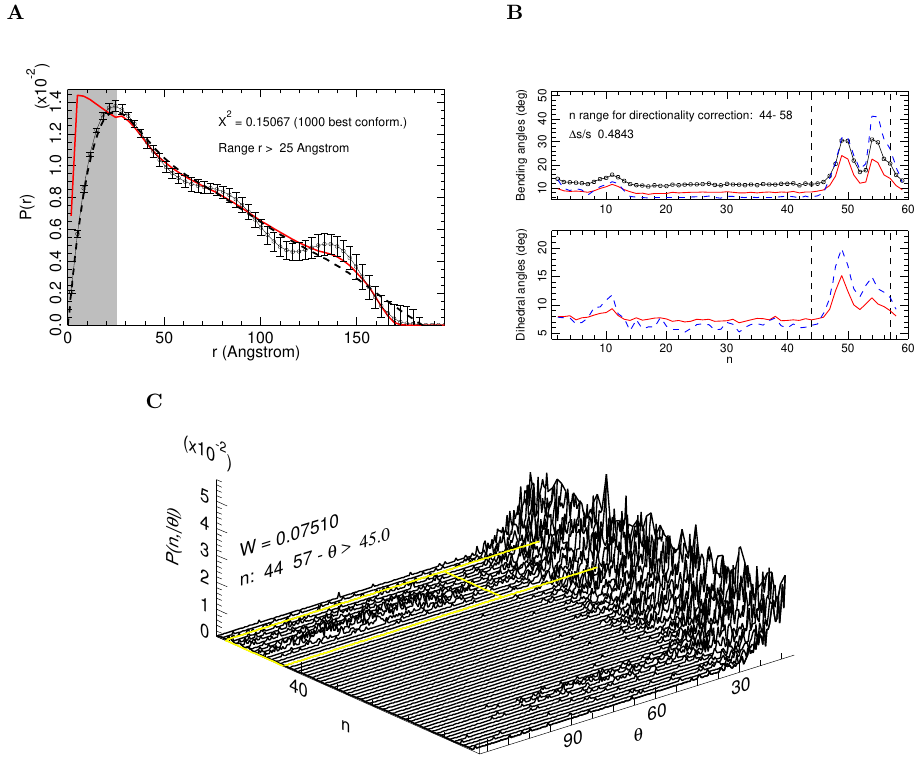}
  \end{center}
  \parbox{17cm}{Figure 8:
    Sequence GAGE6\_2: (A) Comparison between
    $P_{\mathrm{exp}}(r)$ calculated with AutoGNOM-5
    (small circle points and thin black
curve with error bars) and $P(r)$ calculated for the 1000 best conformations
generated theoretically (thick red curve). The grey region, for $r <
25\,$\AA\  is
the low-$r$ domain in which the model cannot describe the internal structure of
the double helix. The thick black dashed line shows the function
$P_{\mathrm{exp}}(r)$ obtained with AutoGNOM-4 (see section 2.4)
(B) Sequence GAGE6\_2:
Upper part: Statistics of the absolute value of the local bending angles
$|\theta_n(i)|$ over the 1000 best-saved conformations for the GAGE6 sample,
after orientation of the conformations as discussed in the Materials and
Methods section. The red full line shows the average over $i$ of
$|\theta_n(i)|$ versus $n$.  The black line with circles shows the same
average, limited to the bending angles which are above $|\theta| = 5^{\circ}$,
and the dashed blue curve shows the standard deviation of $|\theta_n(i)|$.
Lower part: Statistics of the absolute values of the dihedral angles The red
full line shows the average over $i$ of $|\varphi_n(i)|$ versus $n$ and the
the dashed blue curve shows their standard deviation.
(C) Sequence GAGE6\_2: Three-dimensional plot of the
normalised probability distribution function $P(n,|\theta|)$.
The data are not shown
for small bending angles, for which the probabilities are the highest, to
better show the data at larger $\theta$.
The value $W$ marked on the figure shows the
fraction of bending angles $|\theta_n|$ above the value shown
below $W$ within the domain indicated by the two values of $n$. 
  }
  \label{fig:resu-gage62}
\end{figure*}

\bigskip
In GAGE6\_2, in addition to replacing the AT base pairs between positions
$28 \le n \le 31$ with GC pairs, the AT pairs
in positions $38 \le n \le 44$ have now also been
replaced with GC pairs. As expected, the change in properties of the sample
with respect to GAGE6 is even more drastic. Figure 8A shows that
 $P_{\mathrm{exp}}(r)$ has
 sharper lumps, slightly moved towards larger values of $r$.
The error bars given
by GNOM in the calculation of  $P_{\mathrm{exp}}(r)$ from $I(q)$
are larger and the fit of this
pair-distance distribution function by the simple model is less accurate
although it remains within the error bars almost everywhere.

\medskip
Figure 8B shows that the distribution of large bending angles along the
sequence, as well as the domains where the dihedral angles have large standard
deviations, have qualitatively changed. Figure 8B appears to be very
consistent with the sequence of GAGE6\_2. Its largest AT domains are
$10 \le n \le 13$ and a large domain from $n = 50$ to $n = 57$
which is only interrupted by a
single GC pair at position $n = 55$. This agrees well with the small increase of
bending and twist fluctuations seen around $n = 10$ in Fig. 8B and the larger
bending angles and twist fluctuations in the range $44 \le n \le 57$
split into two
peaks by a dip in sites $52,53$ which could correspond to the GC pair at site
$55$ 
within the resolution given by the analysis of the SAXS data. The bending
angles $|\theta_n(i)|$
in this domain are however much smaller than in the sharp
bending domain of GAGE6\_1. This qualitative change from GAGE6\_1 may
nevertheless seem surprising if we think that the only difference in sequence
between GAGE6\_1 and GAGE6\_2 is the elimination of an AT-rich domain between
positions $38 \le n \le 44$.
Both GAGE6\_1 and GAGE6\_2 have a large AT-rich domain
($50 \le n \le 57$) but in GAGE6\_2 this region
causes a smaller bending effect than
GAGE6\_1. This points out again that, although sequence and flexibility are
correlated, the preferred conformations of a DNA sequence in solution are not
only determined by the local sequence but collective effects at larger
distances are also important. The probability
distribution of the bending angles $P(n, |\theta|)$ shown in
Fig. 8C and Supplementary Fig.~S9 are somewhat similar to the data from
Fig. 7:  the fraction of large bending angles $|\theta_n|$ in the range
$44 \le n \le 57$ is of
the order of 9.6\%, suggesting that GAGE6\_2 is also sharply bent in a narrow
region.

\medskip

However, the results shown in Fig. 8A raise questions because the fit of
the polymer model to the $P_{\mathrm{exp}}(r)$ calculated with AutoGNOM-5 looks
rather poor.
We know that the model is oversimplified to fully describe DNA, but, for
$r > 25\;$\AA\ it should nevertheless provide a better description of the
experiment because we have extensively explored the conformational
space of the sample
by generating more than $4\,10^8$ conformations and selecting the best
1000 for the analysis. This failure suggests that the difficulty could come
from $P_{\mathrm{exp}}(r)$ itself. {\em At this point we would like to stress the
importance of choosing a model which has realistic bending and torsional
properties to make sure that the conformations which enter into our fits are
actually accessible to DNA molecules.} A purely random model, such as the
Gaussian model which can adopt any conformation at no energy cost might be
able to provide a good fit to $P_{\mathrm{exp}}(r)$, but with unphysical DNA
conformations.
In section 2.4 we discussed that, for sequence
GAGE6\_2, the determination of the optimal regularisation parameter $\alpha$
by GNOM is particularly hard. GNOM-5 converges to $\alpha = 0.01084$ while
GNOM-4 converges to $\alpha = 1.31$ and generates a much smoother
$P_{\mathrm{exp}}(r)$ with smaller estimated error bars. We conjectured that, for
a given sample, the functions $P_{\mathrm{exp}}(r)$ for different values of
$\alpha$ could contain almost the same information on the physical properties
of the DNA sample. To check this conjecture we repeated with the
 AutoGNOM-4-$P_{\mathrm{exp}}(r)$ functions the same analysis
initially performed with those derived using  AutoGNOM-5. The results are shown
on Fig. 9.

\setcounter{figure}{8}
\begin{figure}
  \centering
    \includegraphics[width=7.5cm]{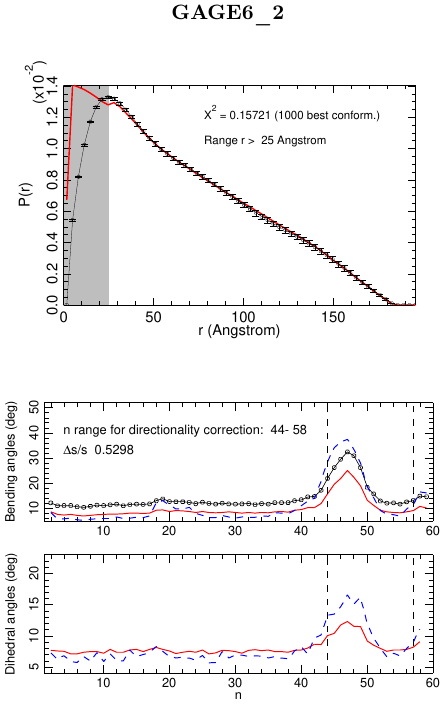}
    \caption{Top panel:
    sequence GAGE6\_2. Comparison of AutoGNOM-4-$P_{\mathrm{exp}}(r)$
    (small circle points and thin black curve with error bars) and $P(r)$
    given by the 1000 best conformations of the polymer model
    (thick red curve). The gray region
    for $r < 25\;$\AA\ is the low-r domain in which the model cannot describe
    the internal structure of the double helix.
    Bottom panel:
    Upper part: Statistics of the absolute value of the local bending angles
$|\theta_n(i)|$ over the 1000 best-saved conformations for the GAGE6 sample,
after orientation of the conformations as discussed in the Materials and
Methods section. The red full line shows the average over $i$ of
$|\theta_n(i)|$ versus $n$.  The black line with circles shows the same
average, limited to the bending angles which are above $|\theta| = 5^{\circ}$,
and the dashed blue curve shows the standard deviation of $|\theta_n(i)|$.
Lower part: Statistics of the absolute values of the dihedral angles The red
full line shows the average over $i$ of $|\varphi_n(i)|$ versus $n$ and the
the dashed blue curve shows their standard deviation.
The vertical dashed lines show the boundaries of the domain where large
bending was observed in the analysis of AutoGNOM-5-$P_{\mathrm{exp}}(r)$.
  }
  \label{fig:GAGE6-2-bis}
\end{figure}

The top panel of Fig.~9  shows that the model gives an almost perfect fit of
AutoGNOM-4-$P_{\mathrm{exp}}(r)$ which is much smoother than
AutoGNOM-5-$P_{\mathrm{exp}}(r)$. In comparing Fig.~8 and Fig.~9, one should keep
in mind that the definition of $\chi^2$ (Eq. (6)) compares the discrepancy
between $P_{\mathrm{exp}}(r)$ and the theoretical $P(r)$ with the estimated error
on $P_{\mathrm{exp}}(r)$. As the error bars on AutoGNOM-4-$P_{\mathrm{exp}}(r)$ are
much smaller than on AutoGNOM-5-$P_{\mathrm{exp}}(r)$ similar values for $\chi^2$
in both figures imply that the discrepancies are much smaller in Fig. 9.
The bottom panel of Fig.~9  shows that, although the two functions
$P_{\mathrm{exp}}(r)$ look 
radically different, they provide almost the same results concerning the
propensity of the DNA sample to bend in the 44-57 domain which is the AT-rich
domain. However, the AutoGNOM-4-$P_{\mathrm{exp}}(r)$ misses the large degree of
bending 
occurring near $n=56-57$ which is detected by the
AutoGNOM-5-$P_{\mathrm{exp}}(r)$. Although it cannot be formally excluded that the
bending detected by AutoGNOM-5-$P_{\mathrm{exp}}(r)$ in the vicinity of an AT
pair is an artefact, it is likely that the high-value of $\alpha$ derived by
AutoGNOM-4 leads to the over-regularisation of this function,
which smoothes out this local
effect. Nevertheless it is remarkable that two functions $P_{\mathrm{exp}}(r)$
which look so different at a first glance encode the same
information about the bendability of a sample near one of its ends. This
supports our conjecture and implies that a perfect refinement of
$P_{\mathrm{exp}}(r)$, which may be difficult, is not an absolute requirement to
analyse SAXS data with a polymer model as proposed in this work. 

\bigskip
In GAGE6\_3, the last of the series of sequences derived from GAGE6, the last
large AT-rich domain that spanned positions $50 \le n \le 54$ has been replaced
entirely by GC base pairs.
Figure 10A shows that $P_{\mathrm{exp}}(r)$
has lost its two-lump structure and conforms in
comparison to the other experimental functions more to the simulated shape of
a straight duplex (Figure 2). The disagreement between the polymer model fit
and $P_{\mathrm{exp}}(r)$ at large values of $r$ here may be a result of the
former not 
accounting for hydration shell effects whilst the latter does.
Although the
large AT-rich regions of the $n > 30$ range that we used to orient the
conformations for previous sequences are no longer present, the sequence still
has asymmetry. The domain $10 \le n \le 26$
has several short AT sequences while its
mirror image $35 \le n \le 49$
now only has three AT pairs and is likely to be less
flexible, being systematically enriched for GC content. This is confirmed by
the sum $s$ and $s'$ of $|\theta_n|$
in these two regions which differ systematically from
each other and can be used to orient the conformations with respect to the
sequence. Figure 10B, which does not display the symmetry expected for
unoriented SAXS data, shows that the orientation procedure described in the
Material and Methods section is again valid for GAGE6\_3.
Figure 10B shows a
broad domain with large bending angles with a sharp maximum at position $n =
26$ 
which coincides with two AT pairs in the middle of a GC sequence. The
three-dimensional plot of the probability 
$P(n,|\theta|)$ plotted in Fig.~10C and
Supplementary Fig.~S9 shows larger bending angles in the domain
$17 \le n \le 35$ but
the integrated weight of the probability $P(n, |\theta|)$ for angles exceeding
$45^{\circ}$ in this
region shows that only 3.8\% of the angles correspond to large bending. As
suggested by its sequence, mostly GC base pairs and only a few small domains
with AT pairs, not wider than three consecutive sites, the GAGE6\_3 sequence
has mostly straight conformations.

\begin{figure*}
  \begin{center}
    \includegraphics[width=16cm]{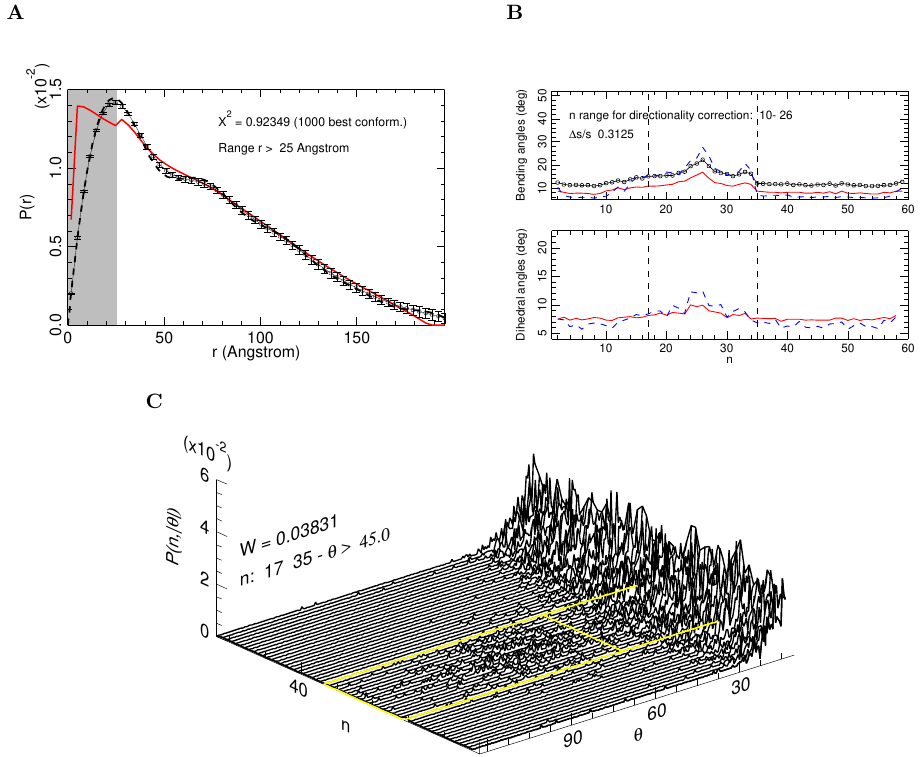}
  \end{center}
  \parbox{17cm}{Figure 10:
    Sequence GAGE6\_3: (A) Comparison between
    $P_{\mathrm{exp}}(r)$ calculated with AutoGNOM-5
    (small circle points and thin black
curve with error bars) and $P(r)$ calculated for the 1000 best conformations
generated theoretically (thick red curve). The grey region, for $r <
25\,$\AA\  is
the low-$r$ domain in which the model cannot describe the internal structure of
the double helix. The thick black dashed line shows the function
$P_{\mathrm{exp}}(r)$ obtained with AutoGNOM-4 (see section 2.4)
(B) Sequence GAGE6\_3:
Upper part: Statistics of the absolute value of the local bending angles
$|\theta_n(i)|$ over the 1000 best-saved conformations for the GAGE6 sample,
after orientation of the conformations as discussed in the Materials and
Methods section. The red full line shows the average over $i$ of
$|\theta_n(i)|$ versus $n$.  The black line with circles shows the same
average, limited to the bending angles which are above $|\theta| = 5^{\circ}$,
and the dashed blue curve shows the standard deviation of $|\theta_n(i)|$.
Lower part: Statistics of the absolute values of the dihedral angles The red
full line shows the average over $i$ of $|\varphi_n(i)|$ versus $n$ and the
the dashed blue curve shows their standard deviation.
(C) Sequence GAGE6\_3: Three-dimensional plot of the
normalised probability distribution function $P(n,|\theta|)$.
The data are not shown
for small bending angles, for which the probabilities are the highest, to
better show the data at larger $\theta$.
The value $W$ marked on the figure shows the
fraction of bending angles $|\theta_n|$ above the value shown
below $W$ within the domain indicated by the two values of $n$. 
  }
  \label{fig:resu-gage63}
\end{figure*}

\medskip

    Supplementary figure S10 shows the results of the
    analysis of the functions $P_{\mathrm{exp}}(r)$ obtained with Auto-GNOM-4
    for the four samples 
    (see supplementary figure Fig.~S6). It can be compared with Figs. 6, 7, 9
    and 10 showing the results obtained with Auto-GNOM-5.
    It confirms our conclusions about the domains prone to extra flexibility
    and bending in each sequence. However, it appears that large regularisation
    parameters $\alpha$, giving smoother $P_{\mathrm{exp}}(r)$, may lead to
    lower amplitudes of the calculated
    bending and flexibility in the DNA sites belonging
    to those domains, and, in some cases loss of fine structures.
    Our results suggest that, for an analysis with a model
    which properly describes DNA bending and torsional rigidity, and therefore
    avoids unphysical conformations, as the choice of the optimal
    regularisation parameter is difficult, favouring  small regularisation
    parameters 
    $\alpha$ may be a good choice because it leads to $P_{\mathrm{exp}}(r)$
    functions which encode more structural details.

\medskip
The study of a series of four DNA sequences, including GAGE6 and three others
obtained by gradually replacing AT-rich domains with GC tracts is interesting
because the unique structural features of bent AT-tract-containing duplexes
can be important for the recognition of dsDNA by proteins \cite{HARTEIS2014}.
Typically, the major groove in dsDNA has the highest
potential for protein-base recognition as the functional groups of the four
bases are accessible \cite{HARTEIS2014}.
In the case of bent
AT-tract-containing structures, the major groove shows an even higher
potential for base recognition as the groove is often widened by bending
\cite{HARTEIS2014,STEFL2004}. Bending likely serves to
enhance sequence-specific recognition of DNA by proteins in many contexts. A
simultaneous effect in AT-tract containing dsDNA is the narrowing of the minor
groove \cite{STEFL2004}. The narrower groove places elements of the
backbone much closer to one another resulting in an enhanced electrostatic
effect that can facilitate structure-specific rather than sequence-specific
recognition by proteins \cite{FERRARI1992,ROHS2009}.

\medskip
It is possible that SFPQ recognises the GAGE6 oligonucleotide through an
interaction of the RGG domains with either a wider major groove or a more
pronounced electrostatic interaction with the backbone due to the narrower
minor groove. The latter is an interesting possibility given that RGG motifs
are the putative DNA-binding regions of SFPQ, and arginine is the most common
residue found to interact with narrow minor grooves \cite{ROHS2009}.
Another possibility is that the fluctuational opening of the “TTTATTT”
region in the GAGE6 oligo, which was the approximate region where the majority
of bending was detected in our analysis, plays a role in protein binding. This
effect may be more pronounced at a higher temperature \cite{DNAFLEX}
or inside biological condensates which through multivalent
competitive interactions can reportedly cause the melting of dsDNA
\cite{NOTT2016}. 
The truncation of the GAGE6 sequence from 60 bp to 40 bp reduced
the binding of SFPQ from 84\% to 45\% as calculated via a gel shift assay
\cite{WANG2022}.
This truncation resulted in the “TTTATTT” domain, being cut in
half. This suggests that the last 20bps of the sequence may provide the region
required for the enhanced binding of SFPQ. However, as functional aggregation
of the protein is a key component of binding \cite{LEE,KONING2025}
it is possible that 40 bp was simply too short a length to allow
multiple units of SFPQ to have several points of contact with the DNA.

\medskip
The idea that certain sites in dsDNA offer less resistance to structural
deformation and so a lower energetic penalty to protein-DNA complexes that
deform rigid DNA is also important for many protein-DNA
interactions. Naturally, if a protein is required to bend DNA as part of its
function, a pre-bent or flexible substrate would dramatically reduce the
amount of energy required to bend DNA. This concept has been reviewed by
\cite{HARTEIS2014} and some examples are:

\textbf{ECOR1} significantly bends DNA at its recognition site ‘GAATTC’ as
part of cleavage reactions \cite{WIDOM1984}.
The degree of natural bending at
this target site likely lowers the energy required for binding as bending of
the unbound DNA can place it close to a position that already resembles the
bound state. Gartenburg and Crothers \cite{GARTENBERG1988} examined
the degree of DNA bending caused by the
\textbf{Catabolite activator protein (CAP)} in
response to different mutated DNA targets. DNA bending in regions flanking the
binding site enhanced the affinity of the CAP-DNA complex by creating a larger
and more complex interface.
\textbf{HMG box domains} in transcriptional regulatory
proteins bind DNA. In general, all of the binding sites for these proteins are
AT-rich, and the protein \textbf{SRY} has been shown
to induce a sharp bend when
binding to the sequence “AACAAG” \cite{FERRARI1992}. It was hypothesised
by Ferrari et al. \cite{FERRARI1992} that this binding site provided a
lower resistance to structural deformation and so lowered the high energetic
cost of bending DNA with an already bent substrate. Nagaich et al.
\cite{NAGAICH1997}
examined the role of \textbf{P53} in DNA bending and deduced that P53 caused
bending in a number of DNA response elements. EMSA assays indicated that
increasingly bent DNA had a higher affinity interaction with P53. Whilst the
examined response elements did not contain strict AT tracts, they did contain
dinucleotide AT regions on either side of the response element.
\textbf{TATA-box
binding protein (TBP)} binding to its DNA target is largely mediated by the
bent shape that its DNA target takes and this contributes to the affinity of
the complex \cite{HARTEIS2014}.
The energetic cost of the DNA’s shape
is lowered by the reduced stacking of the TA steps in the TATA sequence.

\medskip
Realising the pre-bent nature of many of these DNA targets is important in
understanding the affinity and energetics of certain protein-DNA
interactions. Protein recognition of AT-rich DNA can be complicated because of
many interrelated effects. However, DNA bending is reportedly sensitive to
environmental conditions such as temperature, salt, and divalent cations
\cite{HARAN2009}.
Given an initial hypothesis of unwinding, bending or
fluctuational opening of dsDNA playing a role in the affinity of a protein-DNA
complex it may be possible to fine-tune environmental conditions to maximise
such effects and so increase the affinity of protein-DNA interactions for
structural studies. Furthermore, experimental data on DNA bending in response
to sequence composition may also be used to train structure prediction tools
such as successors to Alphafold 3. Our polymer model approach may also perhaps
present an opportunity for the development of other multiphase modelling
systems, that can separately model both the DNA by itself and in complex with
a protein. This could be technically possible via a contrast matching small
angle neutron scattering experiment where an experimentalist could make the
protein “invisible” and only analyse the DNA in an unbound and also
protein-bound state. 

\section{Conclusion}

In this work, we have shown that SAXS data can be analysed to determine the
statistical properties of the conformation of short DNA sequences in
solution. Although the SAXS structure factor averages over all spatial
orientations of the molecules, for sequences which have some asymmetry, even
when it is rather weak such as for GAGE6\_3, the computed conformations can be
oriented with respect to the sequence so that specific features detected by
the analysis can be related to the DNA sequence.

\smallskip
Our analysis uses a polymer model which is simple enough to allow a very broad
exploration of conformational space (up to $10^8$ or $10^9$ conformations with
moderate computing facilities) but is nevertheless able to quantitatively
describe the average persistence length and torsional rigidity of the DNA
double helix.
This is important to ensure that only conformations accessible
  to a DNA molecule are generated. In contrast to the protein chains in
  intrinsically disordered protein regions, which can be described by a
random Gaussian chain \cite{MARTIN2021}, the bending and torsional rigidities
of the double helix introduce strong constraints on the conformational space,
which are intrinsic to our polymer model.
The model, which is restricted to a chain of rigid segments,
does not impose any constraints tying bending to twist. Nevertheless, our
analysis detects the increase of twist fluctuations which is expected for DNA
undergoing strong bending due to the mechanical properties of the double
helix. The ability of the analysis to detect a property of DNA which in itself
is not built into the model strengthens the credibility of the method, as well
as its capability to detect regions which are prone to bending in correlation
with the sequence.

\smallskip
As expected the results confirm that AT-rich regions are more flexible than
GC-rich domains, however, our results have also shown that the link between
sequence and the preferred conformations of free DNA in solution are not
trivially imposed by the local sequence. Using a series of sequences derived
from one another by small local changes we have exhibited unexpected
properties which suggest that collective effects are also very important.

\smallskip
  Our method relies on a fit of $P_{\mathrm{exp}}(r)$, i.e.\ analysis carried out
in real space, rather than handling the data in reciprocal space. This
introduces some difficulty because an accurate derivation of
$P_{\mathrm{exp}}(r)$ 
from the experimental data is not straightforward although it has been the
object of numerous investigations. The determination of the regularisation
parameter $\alpha$ which enters into the calculation is delicate. However,
our analysis has shown  that, when $\alpha$ is modified, the results on the
properties of the sample are robust.

\smallskip

It turns out that  the view in real space is essential to detect {\em local}
effects which are 
spread out in Fourier space. This is not specific to SAXS and spans various
domains of physics. A famous example has been provided by nonlinear localised
waves in water or solid state physics \cite{DAUXOIS2005,SOLITONBOOK}.
A numerical
experiment on vibrational waves to study thermalisation in solids, by Enrico
Fermi and coworkers, raised a puzzling question that was unexplained for more
than 10 years. It stayed a mystery because, being a wave problem, the results
were displayed in the wavevector space ($q$ space). It was only when some
physicists looked in real space that they got a clue because they detected {\em
localised} phenomena (later called solitons) which could easily solve the
puzzle. The real and Fourier spaces bring complementary views, and although
experiments and theory often lead naturally
to the Fourier space, the real space approach may
sometimes be the most appropriate.

\smallskip
This work highlights the interest of SAXS studies to investigate protein
binding sites on DNA because they are able to probe the conformations of DNA
in solution and their dynamics, which is reflected in the statistics of the
conformations provided by SAXS. Crystallographic and CryoEM studies, widely
used to determine the structure of protein-DNA complexes are very precise with
their final structure, in which the conformational freedom of DNA is often
highly constrained by nature of being bound to a protein or vitrification and
crystallisation. However, when proteins scan DNA before binding, they often
sample and test these free-DNA conformations which can be observed by SAXS
studies in solution. Therefore SAXS, crystallographic, and cryoEM studies are
complementary. Having shown that the analysis of SAXS data is able to
determine the properties of DNA in solution at a scale of a few base pairs
enhances the interest in the combination of the two methods to investigate
DNA-protein complexes.  

\bigskip
\noindent\textbf{Acknowledgements}
\ack{Aspects of this research were undertaken on the SAXS/WAXS beamline at the
Australian Synchrotron (Victoria, Australia), part of ANSTO and we thank the
beamline staff for their enthusiastic and professional support.
This work was funded by the Australian Research Council (DP220103667 to
C.S.B.), the National Health and Medical Research Council of Australia
(APP1147496 to C.S.B.) }

\referencelist[saxsdna]

\appendix
\onecolumn
\newpage

\textbf{\Large{Supplementary Materials: }}

  \begin{center}
  \includegraphics[width=13.5cm]{SupplementaryFigure1-redone.png}

\vspace{0.5cm}
\parbox{17cm}{\flushleft
    Supplementary Figure S1: Figure S1 shows the absorbance traces
    of all the 
duplexes at 214nm, 260nm and 280nm vs time (seconds) followed by scattering
traces of all the duplexes plotted from CRHOMIXS showing average X-ray
intensity against frame number. Predicted Rg values are shown as blue curves
and were predicted using CHROMIXS. The regions used for buffer subtraction are
shown in orange and the regions used for sample processing in green. 
}
\end{center}

\newpage

 \begin{center}
  \begin{tabular}{|l|c|c|c|c|}
    \hline
    $\Delta q = 0.001\,$\AA$^{-1}$ & GAGE6~~ & GAGE6\_1 & GAGE6\_2 & GAGE6\_3 \\
    \hline
    GAGE6~~ & - & - & - & - \\
    \hline
    GAGE6\_1 & 0.493 & - & - & - \\
    \hline
    GAGE6\_2 & 0.280 & 0.493 & - & - \\
    \hline
    GAGE6\_3 & 0.078 & 0.493 & 0.753 & - \\
    \hline
  \end{tabular}
  
  \parbox{17cm}{\flushleft
    Supplementary Table 1: Correlation Map Test P-values after
    regridding the data truncated to $q \le 0.18\;$\AA$^{-1}$
    which is the domain that we use in the analysis. }
   \end{center}

  \begin{center}
    \includegraphics[height=15cm]{Vc-MoW-calc.png}
  
\bigskip
  \parbox{17cm}{\flushleft
    Supplementary Figure S2: (A) The GAGE6\_1 mean intensity vs. frame number
chromatogram (pink) shown alongside the results of the MoW calculation
(black squares with light blue error bars) as performed by BioXTAS RAW. The
expected mass of a 60bp dsDNA duplex is shown as a grey dotted horizontal
line. The frames chosen for downstream analysis of P(r) are highlighted by a
light green box. (B) The same as above but for GAGE6\_2. (C) The same as
above but for GAGE6\_3. 
}
\end{center}

\newpage

\begin{center}
     \includegraphics[width=16cm]{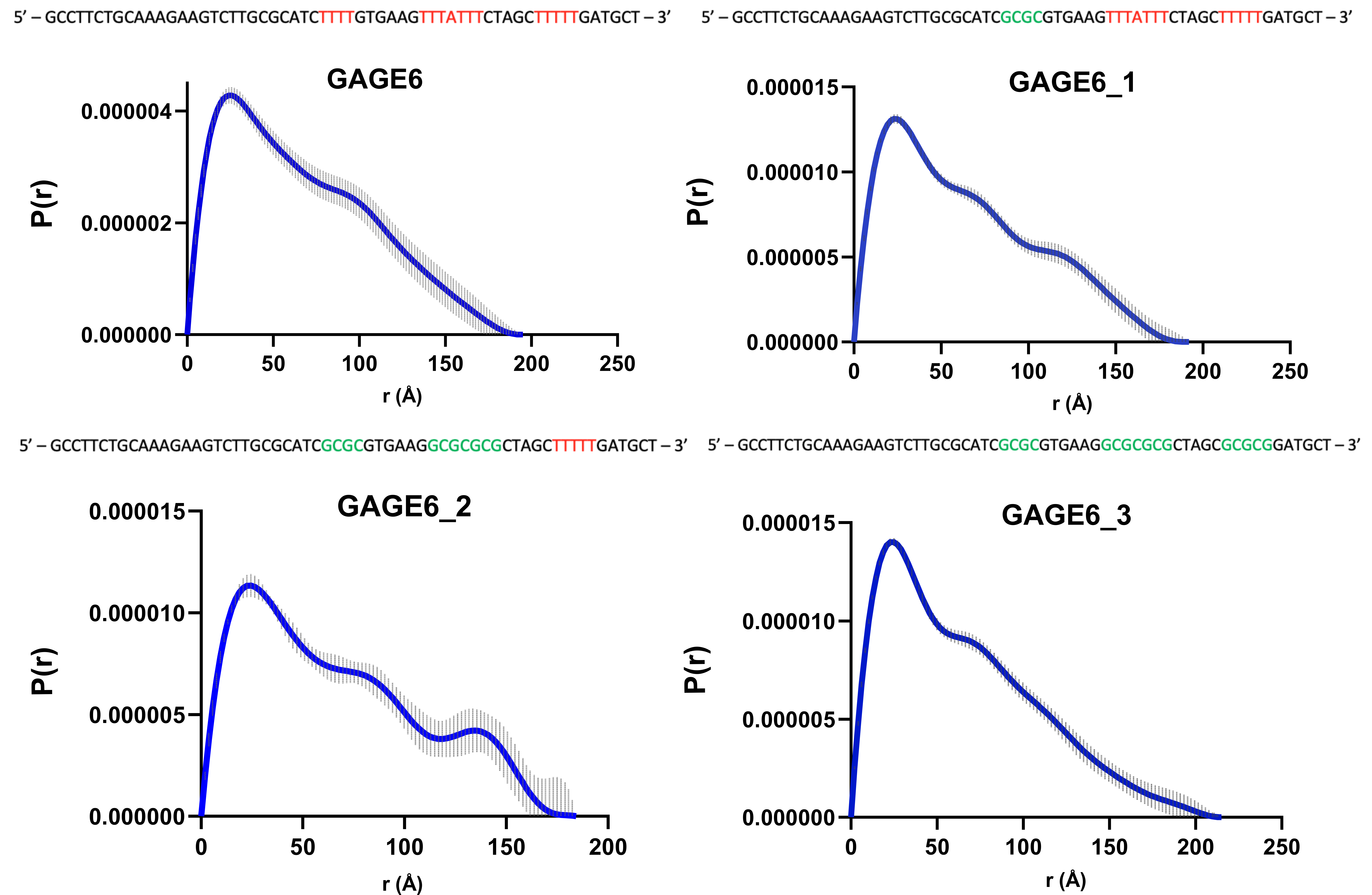}

\bigskip
   
   \parbox{17cm}{\flushleft
     Supplementary Figure S3: Comparative distance distribution
     functions AutoGNOM-5-$P_{\mathrm{exp}}(r)$
     of the different oligonucleotides used in this study. Sequence
    of each oligonucleotide indicated above $P_{\mathrm{exp}}(r)$
    function with AT tracts
    highlighted in red and GC tracts in green. Sequence-dependent changes in
    the $P_{\mathrm{exp}}(r)$
    function indicate the position of multiple bumps in the AT
    tract containing duplexes and a smooth descent for GAGE6\_3.  
  }
\end{center}

\newpage

 \begin{center}
  \includegraphics[width=16cm]{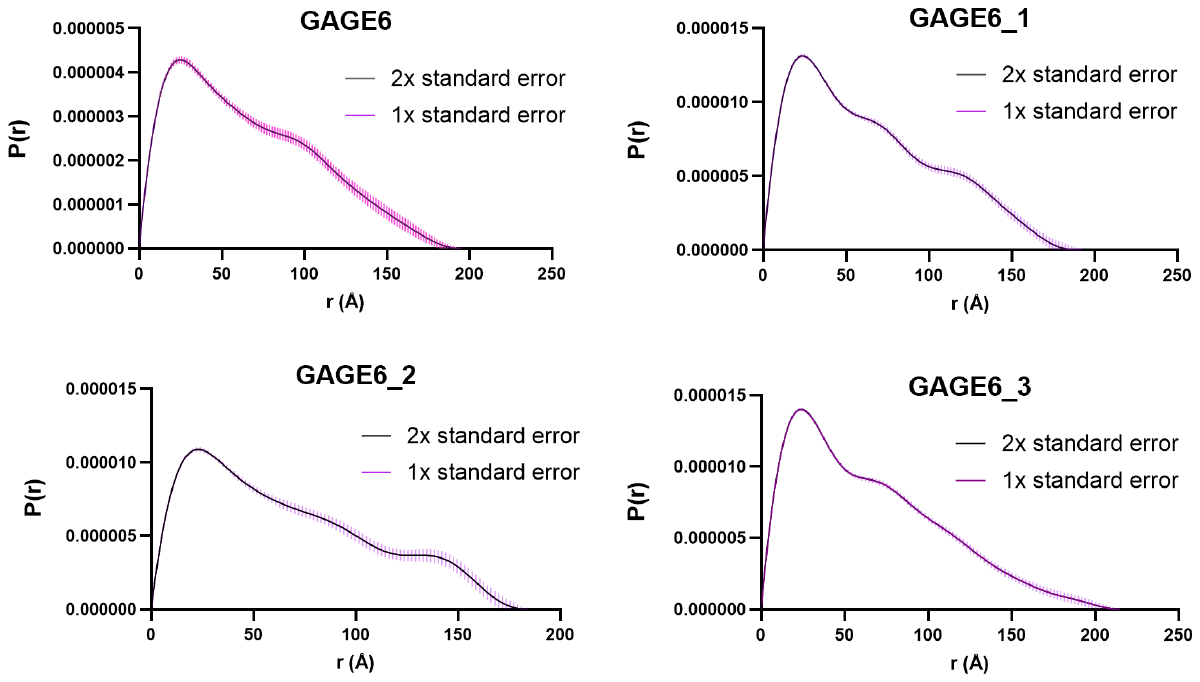}

  \vspace{-13cm}
  \parbox{17cm}{
    \flushleft
    Supplementary Figure S4: Comparative AutoGNOM-5-$P_{\mathrm{exp}}(r)$
    functions for
    datasets accompanied by 1x (pink line) and 2x (black line) standard error
    show consistency in the shape of P(r). Errors bars only plotted for data
    from 1x standard error (pink). 
  }
  \end{center}


  \begin{center}
  \includegraphics[width=16cm]{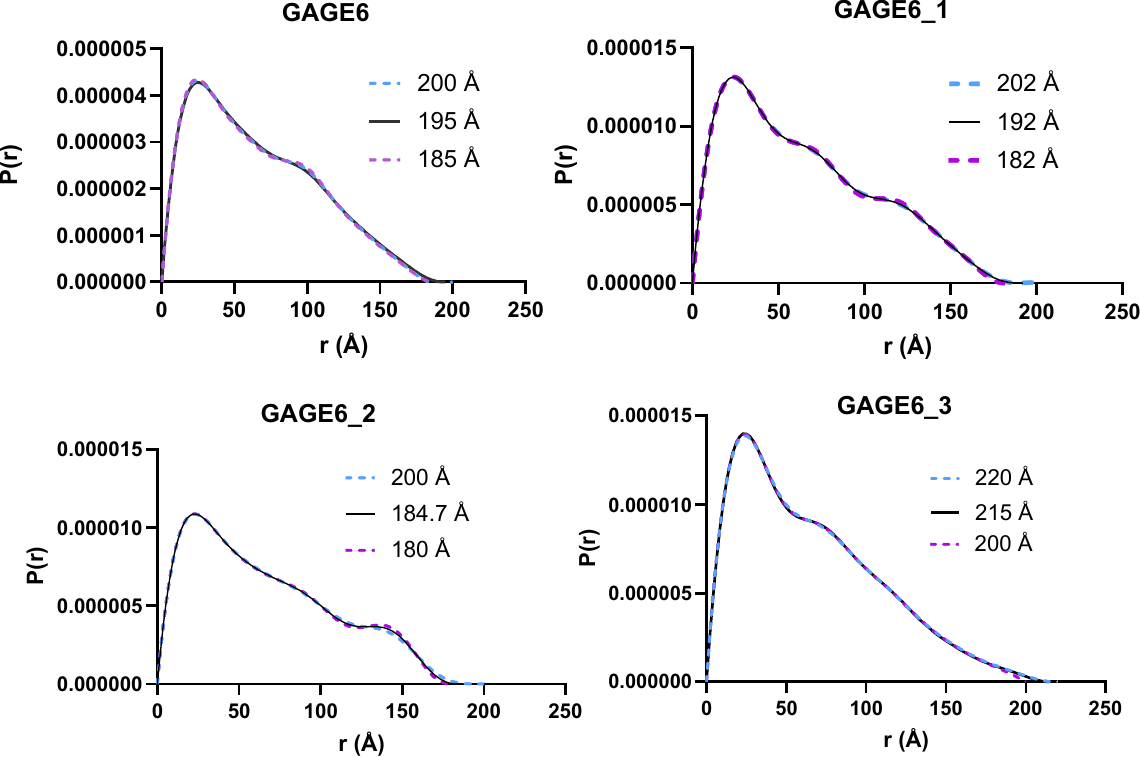}

  \bigskip
  \parbox{17cm}{\flushleft Supplementary Figure S5: Variation of
    $D_{\mathrm{max}}$ parameter around
    the reported values shows that AutoGNOM-5-$P_{\mathrm{exp}}(r)$
    features remain consistently present. 
  }
  \end{center}

  \begin{center}
  \includegraphics[width=16cm]{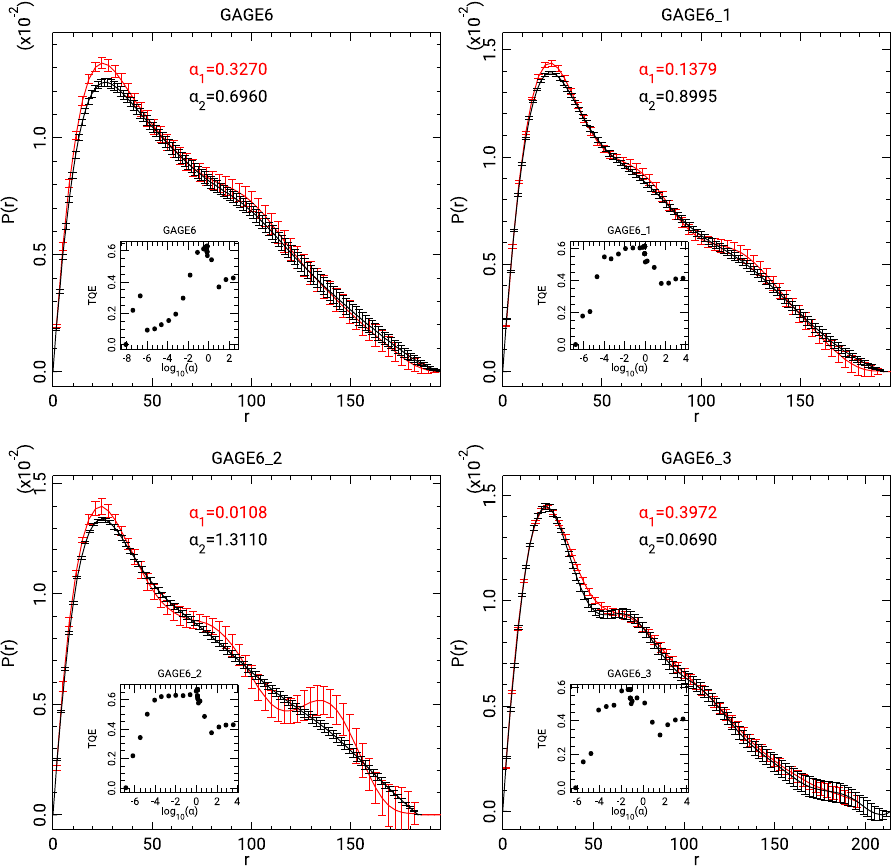}
  
  \parbox{17cm}{\flushleft
    Supplementary Figure S6: Comparison between the functions
    $P_{\mathrm{exp}}(r)$ calculated with AutoGNOM-5 (red curve with error
    bars) and with AutoGNOM-4 (black curve with error bars) for the
    four DNA samples. The values of the
    regularisation parameter $\alpha$ used for each curve are indicated inside
    each panel.
    The inset panels show the variation of the Total Quality Estimate (TQE)
    versus $\alpha$ (in logarithmic scale) calculated by AutoGNOM-4.
  }
  \end{center}

\newpage
  \begin{center}
  \includegraphics[width=16cm]{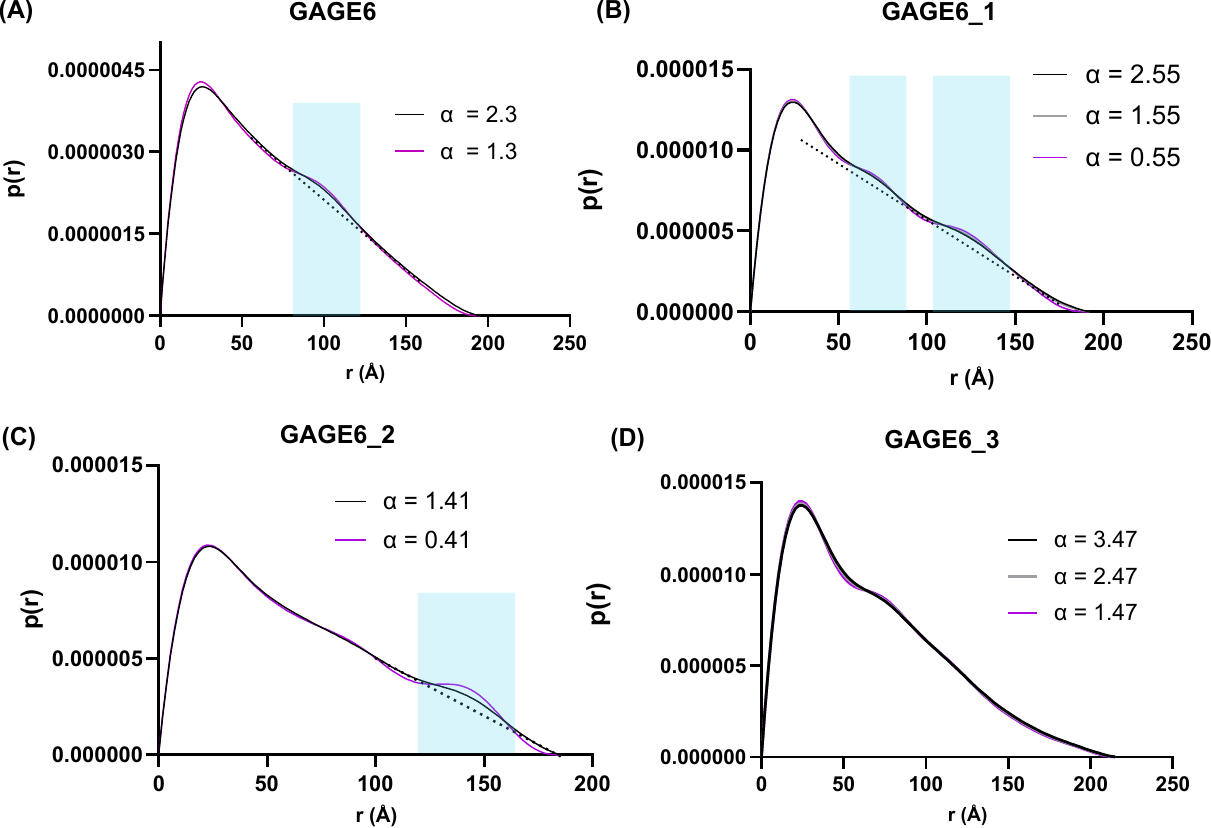}

  \bigskip
  \parbox{17cm}{\flushleft
    Supplementary Figure S7: Variation of the GNOM alpha parameter
    around the fitted value indicates that while the magnitude of features
    changes, their position relative to r remains consistent. 
  }
   \end{center}

\newpage
  \begin{center}

  \includegraphics[width=17cm]{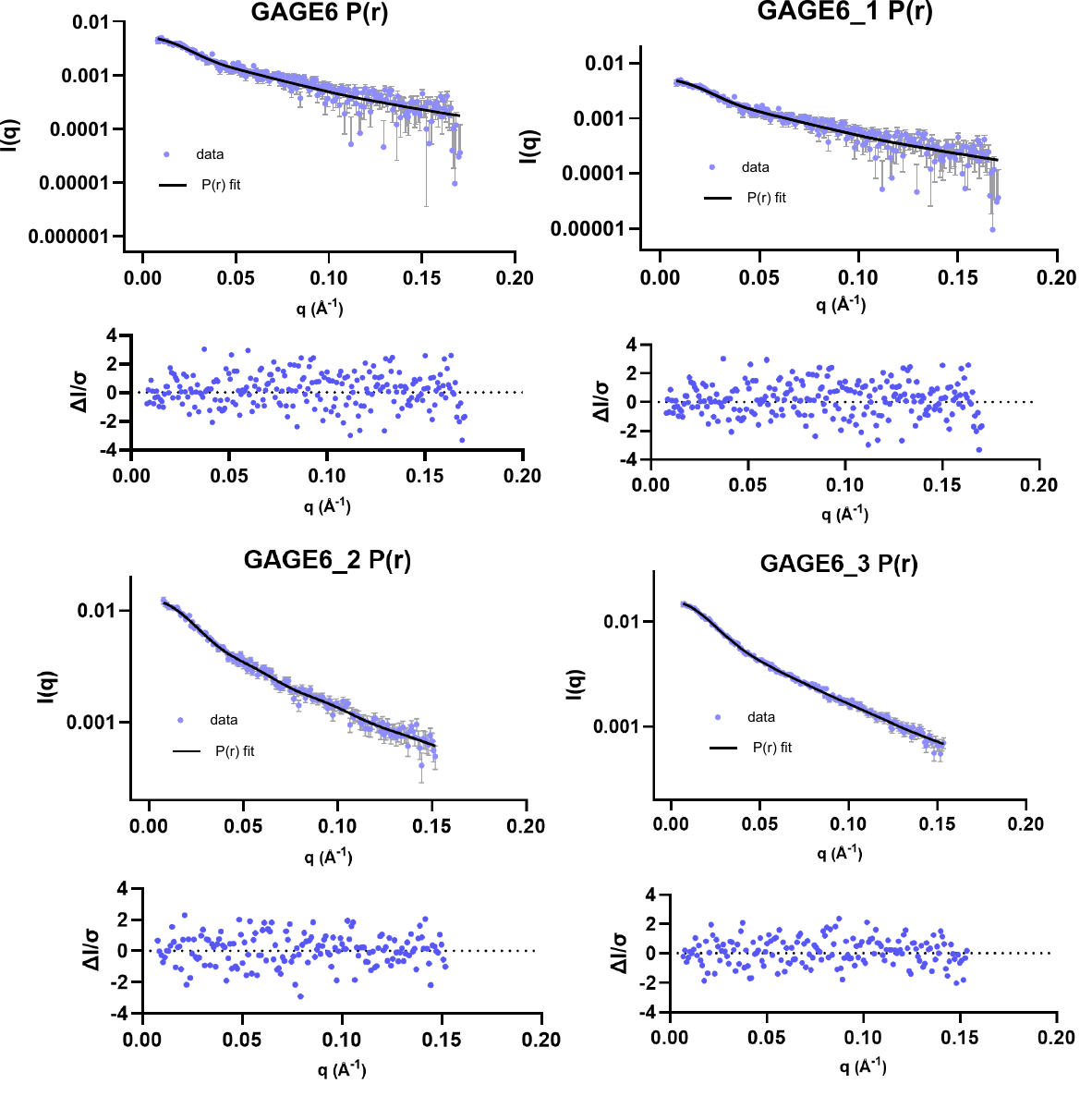}

  \bigskip
  \parbox{17cm}{\flushleft
    Supplementary Figure S8: Plots of the GNOM fits to the
    scattering data and their corresponding residual plots.  
  }
  \end{center}

\newpage
\begin{center}
  \includegraphics[width=17cm]{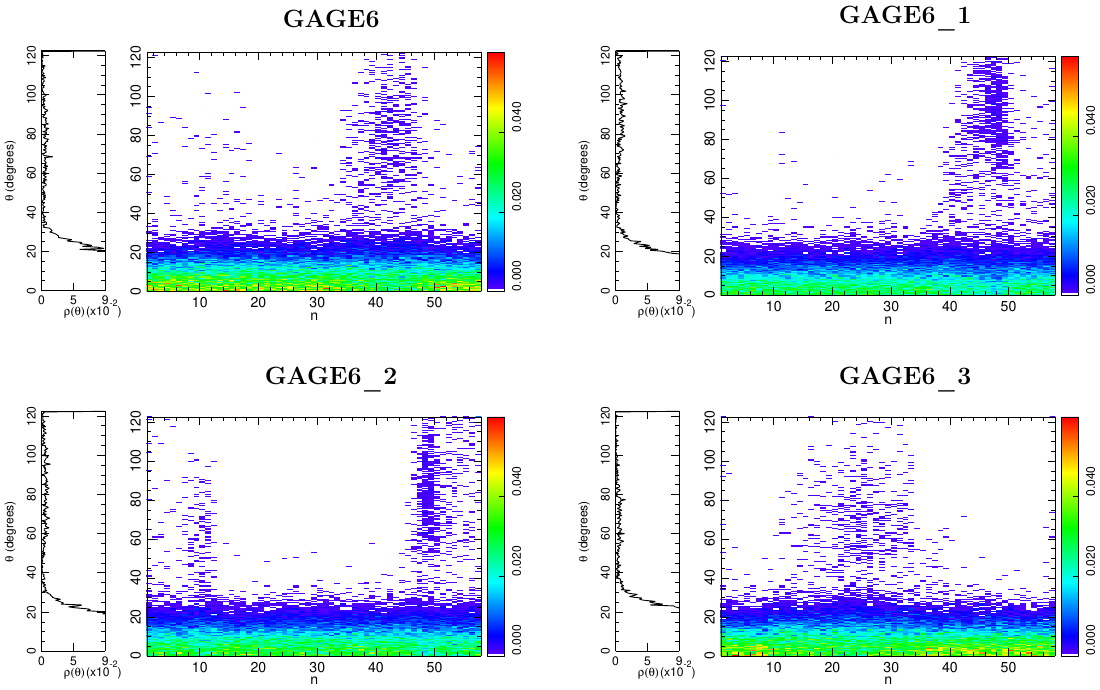}
  
  \parbox{17cm}{\flushleft
    Supplementary Figure S9: Another view of the three-dimensional
probabilities $P(n,|\theta|)$ which highlights some features that are
difficult to see on the three-dimensional plots. The left part shows the sum
over $n$ of the projections of $P(n,|\theta|)$ on a single plane
which displays the probability of a given value of $|\theta|$, whatever the
position in the sequence:

    \begin{equation}
      \label{eq:hsum}
      \rho(\theta) = \sum_{n=2}^{N-1} P(n,|\theta|)
    \end{equation}

  The right part uses a color code to show the three-dimensional probability in
  a two-dimensional image. This is easier to get an idea of the map of the
  bending angles along the sequence. 
}
\end{center}

\newpage
\begin{center}
  \begin{tabular}{cc}
    \includegraphics[width=6.2cm]{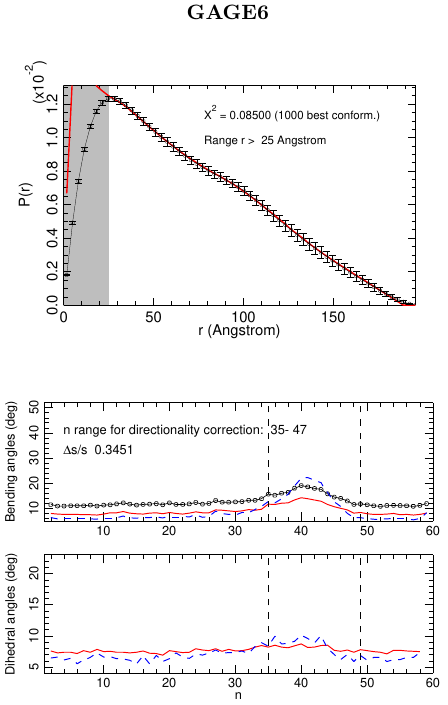} &
    \includegraphics[width=6.2cm]{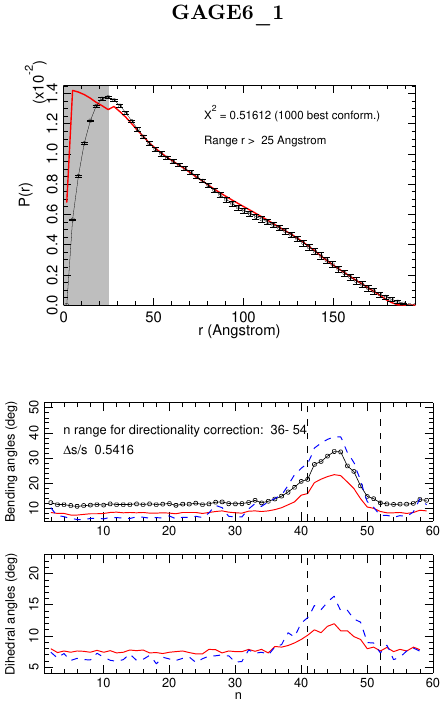} \\
    \includegraphics[width=6.2cm]{figGAGE6_2-compare.pdf} &
     \includegraphics[width=6.2cm]{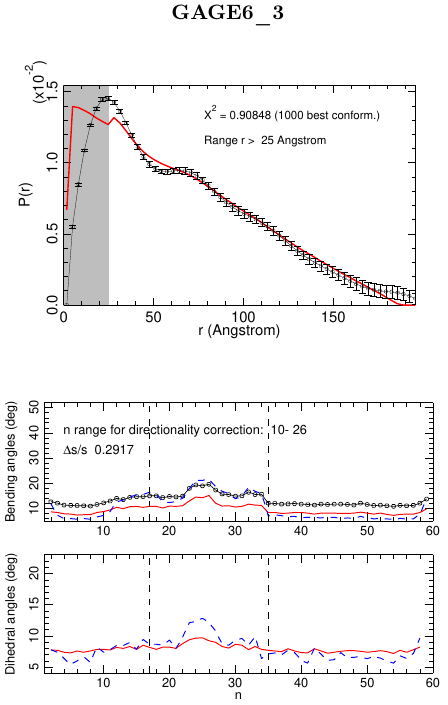} \\
  \end{tabular}

  \parbox{17cm}{\flushleft{
      Supplementary figure S10: Analysis of the $P_{\mathrm{exp}}(r)$ obtained
      with Auto-GNOM-4 (see Fig. S6).
      For each sample the top panel shows the function
      $P_{\mathrm{exp}}(r)$ with error bars (black) and its fit by the polymer
      model 
      (red full line). The bottom panel shows the statistics of the local
      bending and dihedral angles, as  on Figs. 6, 7, 8, 10. The
      vertical dashed lines, limiting the domain showing the largest bending
      are at the same positions as on the corresponding figures of the paper.
      The particular case of GAGE6\_2 is also shown in the paper as Fig.~9.
    }}
\end{center}


@ARTICLE{LOWARY,
  author = "Lowary, P.T.  and Widom, J.",
  title = "New DNA Sequence Rules for High Affinity Binding to Histone Octamer
  and Sequence-directed Nucleosome Positioning",
  journal = "J. Mol. Biol.",
  volume = "276",
  pages = "19--42",
  year = 1998
}

@ARTICLE{HEUMANN,
  author = "Heumann, H. and Richetti, M. and Werel, W.",
  title = "DNA-dependent RNA polymerase of E. coli induces bending or an
   increased flexibility of DNA by specific complex formation",
  journal = "The EMBO Journal",
  volume = "7",
  pages = "4379--4381",
  year = 1988
}

@ARTICLE{KNOTT,
  author = "Knott, G.J. and Bond, C.S. and Fox, A.H.", 
  title = "The DBHS proteins SFPQ, NONO and PSPC1: a multipurpose molecular
   scaffold",
  journal = "Nucleic Acids Research",
  volume = "44",
  pages = "3989--4004",
  year = 2016
}

@ARTICLE{ANSELMI,
  author = "Anselmi, C. and Bocchinfuso, G. and De Santis, P. and 
  Savino M. and Scipioni, A.",
  title = " Dual Role of DNA Intrinsic Curvature and Flexibility in
  Determining Nucleosome Stability",
  journal = "J. Mol. Biol.",
  volume = "286",
  pages = "1293--1301",
  year = 1999
}

@ARTICLE{CROTHERS,
  author = "Crothers, D.M.",
  title = " DNA curvature and deformation in protein?DNA complexes:
  A step in the right direction",
  journal = "Proc. Natl. Acad. Sci. USA",
  volume = "95",
  pages = "15613--15615",
  year = 1988
}

@ARTICLE{DICKERSON,
  author = "Dickerson, R.E. and Drew, H.R.",
  title = ": Structure of a B-DNA Dodecamer. II.
  Influence of Base Sequence on Helix   Structure",
  journal = "J. Mol. Biol.",
  volume = "149",
  pages = "761--786",
  year =  1981
}

@ARTICLE{OLSON1998,
  author = "Olson, W.K. and Gorin, A.A. and Lu, Xiang-Jun and Hock, L. 
  and Zhurkin, V.B.",
  title = " DNA sequence-dependent deformability deduced form protein-DNA 
  crystal complexes",
  journal = "Proc. Natl. Acad. Sci. USA",
  volume = "95",
  pages = "11163--11168",
  year = 1998
}

@ARTICLE{GORIN,
  author = "Gorin, A. A. and Zhurkin, V.B.  and Olson, W. K.",
  title = "B-DNA Twisting Correlates with Base-pair Morphology",
  journal = "J. Mol. Biol.",
  volume = "247",
  pages = "34–48",
  year = 1995
}

@ARTICLE{DNAFLEX,
  author = "Theodorakopoulos, N. and Peyrard, M.",
  title = "Base Pair Openings and Temperature Dependence of DNA Flexibility",
  journal = "PRL",
  volume = "108",
  pages = "078104",
  year = 2012
}

@ARTICLE{LANKAS,
  author = "Lankas F., \v{S}poner J., Langowski J. and Cheatham Ill, T.E.",
  title = "DNA basepair step deformability inferred from molecular dynamics
  simulations",
  journal = "Biophys. J.",
  volume = "85",
  pages = "2872-2883",
  year = 2003
}

@ARTICLE{CURUKSU,
  author = "Curukscu J., Zacharias M., Lavery R. and Zakrzewska, K.",
  title = "Local and global effects of strong DNA bending
  induced during molecular dynamics simulations",
  journal = "Nucleic Acids Research",
  volume = "37",
  pages = " 3766–3773",
  year = 2009
}

@ARTICLE{PETERLIN,
  author = "Peterlin, A.",
  title = "Light scattering by very Stiff Chain Molecules",
  journal = "Nature",
  volume = "171",
  pages = "259--260",
  year = 1953
}

@ARTICLE{LAVERY2010,
  author = "Lavery, R. and others",
  title = "A systematic molecular dynamics study of nearestneighbor
  effects on base pair and base pair step
  conformations and fluctuations in B-DNA",
  journal = "Nucleic Acids Research",
  volume = "38",
  pages = "299--313",
  year = 2010
}

@ARTICLE{SONG,
  author = "Song, X. and Sun, Y. and Garen, A.",
  title = "Roles of PSF protein and VL30 RNA in reversible
   gene regulation",
  journal = "Proc. Natl. Acad. Sci. U.S.A.",
  volume = "102",
  pages = "12189--12193",
  year = 2005
}

@ARTICLE{URBAN,
  author = "Urban, R.J. and Bodenburg, Y.H. and Wood, T.G.",
  title = "NH2 terminus of PTB-associated
    splicing factor binds to the porcine P450scc IGF-I response element",
  journal = "Am. J. Physiol. Endocrinol. Metab.",
  volume = "283",
  pages = "E423--427",
  year = 2002
}

@ARTICLE{LEE,
  author = "Lee, M. and others",
  title = "The structure of human SFPQ reveals a coiled-coil mediated
    polymer essential for functional aggregation in gene regulation",
  journal = "Nucleic Acids Res.",
  volume = "43",
  pages = "3826--3840",
  year = 2015
}

@ARTICLE{WANG2022,
  author = "Wang, J. and others",
  title = "Insight into the Tumor Suppression Mechanism from the
  Structure of Human Polypyrimidine Splicing Factor (PSF/SFPQ) Complexed with
  a 30mer RNA from Murine Virus-like 30S Transcript-1",
  journal = "Biochemistry",
  volume = "61",
  pages = "1723--1734",
  year = 2022
}

@ARTICLE{MPKINK,
  author = "Schindler, T. and others",
  title = " Kinky DNA in solution: Small angle
  scattering study of a nucleosome positioning sequence",
  journal = "Phys. Rev. E",
  volume = "98",
  pages = "042417",
  year = 2018
}

@ARTICLE{CHONG,
  author = "Chong, P. A. and Vernon, R. M. and Forman-Kay, J. D.",
  title = "RGG/RG Motif Regions in RNA Binding and Phase Separation",
  journal = "J. Mol. Biol.",
  volume ="430", 
  pages = "4650–-4665",
  year = 2018
}

@ARTICLE{KIRBY,
  author = "Kirby, N. and others",
  title = "Improved radiation dose efficiency in solution SAXS using a
    sheath flow sample environment",
  journal = "Acta Crystallogr. D Struct. Biol.",
  volume = "72", 
  pages = "1254–1266",
  year = 2016
}

@ARTICLE{RYAN,
  author = "Ryan, T. M. and others",
  title = "An optimized SEC-SAXS system enabling high X-ray dose for
  rapid SAXS assessment with correlated UV measurements for biomolecular
  structure analysis", 
  journal = "J. Appl. Crystallogr.",
  volume = "51",
  pages = "97–-111",
  year = 2018
}

@ARTICLE{PETOUKHOV,
  author = "Petoukhov, M. V. and Franke, D. and Shkumatov, A.V.
  and Tria, G. and Kikhney, A.G. and Gajda, M. and Gorba, C. and
  Mertens, H.D.T. and  Konarev, P.V. and Svergun, D.I.",
  title = "New developments in the ATSAS program package for
    small-angle scattering data analysis",
  journal = "J. Appl. Crystallogr.",
  volume = "45",
  pages = "342–-350",
  year = 2012
}

@ARTICLE{TREWHELLA,
  author = "Trewhella, J. and  Duff, A.P. and  Durand, D.
 and  Gagel, F. and Guss, J.M. and 
 Hendrickson, W.A. and Hura, G.L. and 
 Jacques, D.A. and Kirby, N.M. and Kwan, A.H. and others",
  title = " ",
  journal = "Acta Crystallogr. D Struct. Biol.",
  volume = "73",
  pages = "710-728",
  year = 2017
}

@ARTICLE{MANALASTAS,
  author = "Manalastas-Cantos, K. and  Konarev, P.V. and  Hajizadeh, N.R. 
  and  Kikhney, A.G. and 
  Petoukhov, M.V. and  Molodensky, D.S. and  Panjkovich, A. 
  and  Mertens, H.D.T. and  Gruzinov, A. and  Borges, C. and others",
  title = " ",
  journal = "J. Appl. Crystallogr.",
  volume = "54",
  pages = "343-355",
  year = 2021
}

@BOOK{SAENGER,
  author = "Saenger, W.",
  title = "Principles of nucleic acid structure",
  publisher = "Springer-Verlag, New York Inc.",
  year = 1984
}

@BOOK{SIVIA,
  author = "Sivia, D.S..",
  title = "Elementary Scattering Theory. For X-ray and Neutron Users",
  publisher = "Oxford University Press Inc. New York",
  year = 2011
}

@ARTICLE{BERNARDO,
 author = "Bernado, P. and Mylonas, E. and Petoukhov, M.V. and
  Blackledge, M. and Svergun, D.I.",
  title = "Structural Characterization of Flexible Proteins Using
  Small-Angle X-ray Scattering",
  journal = "J. Am. Chem. Soc.",
  volume = "129",
  pages = "5656--5664",
  year = 2007
}

@ARTICLE{SCHELLMAN1974,
  author = "Schellman, J.A.",
  title = "Flexibility of DNA",
  journal = "Biopolymers",
  volume = "13",
  pages = "217-226",
  year = 1974
}

@ARTICLE{SMITH-SB92,
  author = "Smith, S.B. and Finzi, L. and C. Bustamante, C.",
  title = "Direct Mechanical Measurements of the Elasticity of Single 
  DNA Molecules Using Magnetic Bead",
  journal = "Science",
  volume = "258",
  pages = "1122--1126",
  year = 1992
}

@ARTICLE{LU-Y,
  author = "Lu, Y. and Weers, B. and Stellwagen, N.C.",
  title = "DNA Persistence Length Revisited",
  journal = "Biopolymers",
  volume = "61",
  pages = "261--275",
  year = 2002
}

@ARTICLE{BRYANT,
  author = "Bryant, Z. and Stone, M.D. and Gore, J. and Smith, S.B.
  and Cozzrelli N.R. and Bustamante, C.",
  title = " Structural transitions and elasticity from torque 
  measurements on DNA",
  journal = "Nature",
  volume = "424",
  pages = "338--341",
  year = 2003
}

@ARTICLE{METROPOLIS1953,
  author = "Metropolis, N. and Rosenbluth, A.W. and Rosenbluth, M.N.
  and Teller, A.H. and Teller, E.",
  title = "Equation of State Calculations by Fast Computing Machines",
  journal = "J. Chem. Phys.",
  volume = "21",
  pages = "1087--1092",
  year = 1953
}

@ARTICLE{WIDOM1984,
  author = "Widom, J.",
  title = "DNA bending and kinking",
  journal = "Nature",
  volume = "309",
  pages = "312--313",
  year = 1984
}

@ARTICLE{STEFL2004,
  author = "Stefl, R. and Wu, H. and Ravindranathan, S. and Sklen\'a\v{r}, V.
    and Feigon, J.",
  title = "DNA A-tract bending in three dimensions:
    Solving the $dA_4T_4$ vs. $dT_4A_4$ conundrum",
  journal = "PNAS",
  volume = "101",
  pages = "1177–-1182",
  year = 2004
}

@ARTICLE{HIZVER2001,
  author = "Hizver, J. and Rozenberg, H. and Frolow, F. and Rabinovich, D.
   and Shakked, Z.",
  title = "DNA bending by an adenine–thymine tract and its
    role in gene regulation",
  journal = "PNAS",
  volume = "98",
  pages = "8490--8495",
  year = 2001
}

@ARTICLE{CHIRICO2001,
  author = "Chirico, G. and Collini, M. and Toth, K. and Brun, N. 
  and Langowski, J.",
  title = "Rotational dynamics of curved DNA fragments studied by fluorescence
    polarization anisotropy",
  journal = "Eur. Biophys. J.",
  volume = "29",
  pages = "597--606",
  year = 2001
}

@ARTICLE{HARTEIS2014,
  author = "Harteis, S. and Schneider, S.",
  title = "Making the Bend: DNA Tertiary Structure and Protein-DNA 
  Interactions",
  journal = "Int. J. Mol. Sci.",
  volume = "15",
  pages = "12335--12363",
  year = 2014
}

@ARTICLE{GUERON1987,
  author = "Gu{\'{e}}ron, M. and Kochoyan, M. and Leroy, J.-L.",
  title = "A single mode of DNA base-pair opening drives imino
    proton exchange",
  journal = "Nature",
  volume = "328",
  pages = "89--92",
  year = 1987
}

@ARTICLE{CHANDRASEKAR2024,
  author = "Chandrasekhar, S. and others",
  title = "Bending Unwinds DNA",
  journal = "{bioRxiv} 2024.02.14.579968",
  howpublished = "doi:10.1101/2024.02.14.579968",
  year = 2024
}

@ARTICLE{FERRARI1992,
  author = "Ferrari, S. and others",
  title = "SRY, like HMG1, recognizes sharp angles in DNA",
  journal = "EMBO J.",
  volume = "11",
  pages = "4497-–4506",
  year = 1992
}

@ARTICLE{ROHS2009,
  author = "Rohs, R. and others",
  title = "The role of DNA shape in protein-DNA recognition",
  journal = "Nature",
  volume = "461",
  pages = "1248-–1253",
  year = 2009
}

@ARTICLE{NOTT2016,
  author = "Nott, T. J. and Craggs, T. D. and Baldwin, A. J.",
  title = "Membraneless organelles can melt nucleic acid duplexes and act as
    biomolecular filters",
  journal = "Nat. Chem.",
  volume = "8",
  pages = "569-–575",
  year = 2016
}

@ARTICLE{GARTENBERG1988,
  author = "Gartenberg, M. R. and Crothers, D. M.",
  title = "DNA sequence determinants of CAP-induced bending and protein binding
    affinity",
  journal = "Nature",
  volume = "333",
  pages = "824-–829",
  year = 1988
}

@ARTICLE{NAGAICH1997,
  author = "Nagaich, A. K. and Appella, E. and Harrington, R. E.",
  title = "DNA bending is essential for the site-specific recognition of DNA
    response elements by the DNA binding domain of the tumor suppressor
    protein p53",
  journal = "J. Biol. Chem.",
  volume = "272",
  pages = "14842–-14849",
  year = 1997
}

@ARTICLE{HARAN2009,
  author = "Haran, T. E. and Mohanty, U.",
  title = "The unique structure of A-tracts and intrinsic DNA bending",
  journal = "Quart. Rev. Biophys.",
  volume = "42",
  pages = "41–81",
  year = 2009
}

@ARTICLE{SVERGUN1992,
  author = "Svergun, D.I.",
  title = "Svergun, A.V. Semenyuk and L. Feigin,
  Small-Angle-Scattering-Data Treatment by the Regularization Method",
  journal = "J. Appl. Cryst.",
  volume = "25",
  pages = "495-503",
  year = 1992
}

@ARTICLE{SVERGUN995,
  author = "Svergun, D.I. and Barberato, C. and Koch, M.",
  title = " ",
  journal = "J. Appl. Cryst.",
  volume = "28",
  pages = "768-773",
  year = 1995
}

@ARTICLE{ABRAMSOL,
  author = "Abramsol, J. and others",
  title = " ",
  journal = "Nature",
  volume = "630",
  pages = "493-500",
  year = 2024
}

@ARTICLE{GLATTER1977,
  author = "Glatter, O.",
  title = "A New Method for the Evaluation of Small-Angle Scattering Data",
  journal = "J. Appl. Cryst.",
  volume = "10",
  pages = "415-421",
  year = 1977
}

@ARTICLE{BLANCHET,
  author = "Blanchet, C.E. and Svergun, D.I.",
  title = "Small-Angle X-Ray Scattering on Biological Macromolecules
  and Nanocomposites in Solution",
  journal = "Annu. Rev. Phys. Chem.",
  volume = "64",
  pages = "37-54",
  year = 2013
}

@ARTICLE{SCHNEIDMAN,
  author = "Schneidman-Duhovny, D. and Hammel, M. and Tainer, J.A. and 
  Sali, A.",
  title = "",
  journal = "Nucleic Acid Res.",
  volume = "44",
  pages = "W424-W429",
  year = 2016
}

@ARTICLE{TRIA,
  author = "Tria, G. and Mertens, H.D.T. and Kachala, M. and Svergun, D.I.",
  title = "Advanced ensemble modelling of flexible macromolecules using X-ray 
  solution scattering", 
  journal = "IUCrj",
  volume = "2",
  pages = "207-217",
  year = 2015
}

@ARTICLE{KONING2025,
  author = "Koning, H.J. and others",
  title = "Structural plasticity of the coiled–coil interactions in 
  human SFPQ", 
  journal = "Nucleic Acids Res.",
  volume = "53",
  pages = "gkae1198",
  year = 2025
}

@ARTICLE{MARTIN2021,
  author = "Martin, E.W. and Hopkins, J.B. and Mittag, T.",
  title = "Small angle x-ray scattering experiments of monodisperse
  intrinsically disordered protein samples close to the solubility
  limit", 
  journal = "Methods Enzymol.",
  volume = "646",
  pages = "185-222",
  year = 2021
}

@ARTICLE{DAUXOIS2005,
  author = "Dauxois, T. and Peyrard, M. and Ruffo, S.",
  title = "The Fermi–Pasta–Ulam ‘numerical experiment’: history and pedagogical
  perspectives.", 
  journal = "Eur. J. Phys.",
  volume = "26",
  pages = "S3-S11",
  year = 2005
}

@BOOK{SOLITONBOOK,
  author = "Dauxois, T. and Peyrard, M.",
  title = "Physics of Solitons (Chap. 1 and 8)",
  publisher = "Cambridge University Press",
  year = 2000
}

@ARTICLE{GRANT2015,
  author = "Grant, T.D. and Luft, J.R. and Carter, L.G. and
  Matsui, T. and Weiss, T.M. and Martel, A. and Snel, E.W.",
  title = "The accurate assessment of small-angle X-ray
  scattering data", 
  journal = "Acta Crystallogr. D Struct. Biol.",
  volume = "71",
  pages = "45-56",
  year = 2015
}

@ARTICLE{FRANKE2015,
  author = "Franke, D. and Jeffries, C.M. and Svergun, D.I.",
  title = "Correlation Map, a goodness-of-fit test for one-dimensional 
  X-ray scattering spectra",  
  journal = "Nat. Methods",
  volume = "12",
  pages = "419-422",
  year = 2015
}

@ARTICLE{CHROMIXS,
  author = "Panjkovich, A. and and Svergun, D.I.",
  title = "CHROMIXS: automatic and interactive analysis of 
  chromatography-coupled small-angle X-ray scattering data",  
  journal = " Bioinformatics",
  volume = "34",
  pages = "1944-1946",
  year = 2018
}

@ARTICLE{HOPKINS2017,
  author = "Hopkins, J.B. and Gillilan, R.E. and Skou, S.",
  title = "BioXTAS RAW: improvements to a free open-source 
  program for small-angle X-ray scattering data reduction and analysis",  
  journal = "J. Appl. Crystallogr.",
  volume = "50",
  pages = "1545-1553",
  year = 2017
}

@ARTICLE{RAMBO2013,
  author = "Rambo, R.P. and Tainer, J.A.",
  title = "Accurate assessment of mass, models and resolution by
small-angle scattering",  
  journal = "Nature",
  volume = "496",
  pages = "477-81",
  doi={https://doi.org/10.1038/nature12070},
  year = 2013
}
\end{document}